\documentstyle[12pt,epsf]{article}
 
\newcommand{\nsect}{\setcounter{equation}{0}
\def\theequation{\thesection.\arabic{equation}}\section}
\newcommand{\nappend}{\setcounter{equation}{0}
\def\theequation{\rm{A}.\arabic{equation}}\section*}

\newcommand{\newc}{\newcommand}

\newc{\del}{\partial}
\newc{\beq}{\begin{equation}}
\newc{\eeq}{\end{equation}}
\newc{\barr}{\begin{eqnarray}}
\newc{\earr}{\end{eqnarray}}

\newc{\ra}{\rightarrow}
\newc{\lam}{\lambda}
\newc{\eps}{\epsilon}
\newc{\half}{\frac{1}{2}}
\newc{\third}{\frac{1}{3}}
\newc{\fourth}{\frac{1}{4}}
\newc{\eighth}{\frac{1}{8}}
\newc{\gev}{\mbox{~GeV}}
\newc{\lra}{\leftrightarrow}
\newc{\Dslash}{\not\!\! D}
\newc{\sg}{{\cal G}}
\newc{\etal}{{\it et al.}\ }
\newc{\Hbar}{{\bar H}}
\newc{\hhbar}{{\overline h}}
\newc{\Ubar}{{\bar U}}
\newc{\Dbar}{{\bar D}}
\newc{\Ebar}{{\bar E}}
\newc{\eg}{{\it e.g.}\ }
\newc{\ie}{{\it i.e.}\ }
\newc{\nonum}{\nonumber}
\newc{\kap}{\kappa}
\newc{\Dt}{\frac{d}{dt}}
\newc{\rpv}{$\not\!\!R_p$}
\newc{\bpv}{$\not\!\!B_p$}
\newc{\mpl}{$M_{Pl}$\ }
\newc{\ol}{\overline}
\newc{\mx}{$M_X$\ }
\newc{\tev}{\mbox{~TeV}}
\newc{\sect}[1]{\ref{sec:#1}}
\newc{\nonr}{\nonumber}
\newc{\eq}[1]{(\ref{eq:#1})}
\newc{\eqs}[2]{(\ref{eq:#1},\ref{eq:#2})}
\newc{\lab}[1]{\label{eq:#1}}
\newc{\Lam}{{\bf\Large \Lambda}}
\newc{\ltau}{\lambda_\tau}
\newc{\lt}{\lambda_t}
\newc{\lb}{\lambda_b}
\newc{\lae}{{\Lam}_E}
\newc{\lad}{{\Lam}_D}
\newc{\lau}{{\Lam}_U}
\newc{\lame}[1]{{\Lam}_{E^{#1}}}
\newc{\lamd}[1]{{\Lam}_{D^{#1}}}
\newc{\lamu}[1]{{\Lam}_{U^{#1}}}
\newc{\lamet}[1]{{\Lam}_{E^{#1}}^T}
\newc{\lamdt}[1]{{\Lam}_{D^{#1}}^T}
\newc{\lamut}[1]{{\Lam}_{U^{#1}}^T}
\newc{\lames}[1]{{\Lam}_{E^{#1}}^*}
\newc{\lamds}[1]{{\Lam}_{D^{#1}}^*}
\newc{\lamus}[1]{{\Lam}_{U^{#1}}^*}
\newc{\lamed}[1]{{\Lam}_{E^{#1}}^\dagg}
\newc{\lamdd}[1]{{\Lam}_{D^{#1}}^\dagg}
\newc{\lamud}[1]{{\Lam}_{U^{#1}}^\dagg}
\newc{\Y}{{\bf Y}}
\newc{\ye}{{\Y}_E}
\newc{\yd}{{\Y}_D}
\newc{\yu}{{\Y}_U}
\newc{\yes}{{\Y}_E^*}
\newc{\yds}{{\Y}_D^*}
\newc{\yus}{{\Y}_U^*}
\newc{\yet}{{\Y}_E^T}
\newc{\ydt}{{\Y}_D^T}
\newc{\yut}{{\Y}_U^T}
\newc{\yed}{{\Y}_E^\dagg}
\newc{\ydd}{{\Y}_D^\dagg}
\newc{\yud}{{\Y}_U^\dagg}
\newc{\dagg}{\dagger}
\newc{\lp}{\left(}
\newc{\rp}{\right)}
\newc{\inv}{\frac{1}{16\pi^2}}
\newc{\invsq}{\frac{1}{(16\pi^2)^2}}
\newc{\ggam}[2]{\Gamma_{#2}^{#1}}
\newc{\yukgam}[2]{\inv \gamma_{#1}^{(1){#2}}+\invsq\gamma_{{#1}}^{(2){#2}}}
\newc{\susyunif}{ohman,nirpaul,marcelacarlos,susyunif}
\newc{\lsim}{\stackrel{<}{\sim}}
\newc{\gsim}{\stackrel{>}{\sim}}
\begin{document}

\title{2-Loop Supersymmetric Renormalisation Group Equations
Including R-Parity Violation and Aspects of Unification}
\author{B.C. Allanach$^{1,a}$, A. Dedes$^{2,b}$, H.K. Dreiner$^{2,c}$}
\date{$^1${\small DAMTP, Silver st, Cambridge CB3
9EW, UK.}\\
$^2${\small Rutherford Appleton Laboratory, Chilton, Didcot, Oxon, OX11
0QX, UK.}}
\maketitle

\vspace{-10.0cm}
\hfill\parbox{8cm}{\raggedleft \today \\ DAMTP-1999-20 \\ hep-ph/9902251}
\vspace{8.5cm}

\begin{abstract}
\noindent We present the complete 2-loop renormalisation group equations of
the superpotential parameters for the supersymmetric standard model
including the full set of $R$-parity violating couplings. We use these
equations to do a study of (a) gauge coupling unification, (b)
bottom-tau unification, (c) the fixed-point structure of the top quark
Yukawa coupling, and (d) two-loop bounds from perturbative unification.  
For large values of the R-parity violating coupling, the value of
$\alpha_S(M_Z)$ predicted from unification can be reduced by 5$\%$
with respect to the R-parity conserving case, bringing it to within
2$\sigma$ of the observed value. Bottom-tau Yukawa unification becomes
potentially valid for any value of $\tan\beta\sim 2-50$. The
prediction of the top Yukawa coupling from the low $\tan\beta$,
infra-red quasi fixed point can be lowered by up to $10\%$, raising
$\tan\beta$ up to a maximum of 5 and relaxing experimental
constraints upon the quasi-fixed point scenario. For heavy scalar
fermion masses ${\mathcal{O}}(1\tev)$ the limits on the higher family
$\Delta L\not=0$ operators from perturbative unification are
competitive with the indirect laboratory bounds. We calculate the
dependence of these bounds upon $\tan \beta$.
\end{abstract}
\vfill
{\tt \begin{flushleft}
a) b.c.allanach@damtp.cam.ac.uk\\
b) adedes@v2.rl.ac.uk\\
c) dreiner@v2.rl.ac.uk\\
\end{flushleft}}

\nsect{Introduction} 
The first grand unified theory (GUT), of the electroweak and the
strong interactions was non-supersymmetric and the unification scale
was of order $M_X=10^{15}\gev$ \cite{georgi}. In order to connect the
GUT predictions at $M_X$ with observations at presently accessible
energies, the renormalisation group evolution of the relevant
parameters must be taken into account \cite{quinn}. Postulating
unification of the gauge couplings at a high scale leads after
renormalisation to one low-energy prediction, \eg the electroweak
mixing angle $\sin^2\theta_W$. In 1987, it was first found that in
supersymmetry the prediction for $\sin^2\theta_W$ is in agreement with
the data, while in the Standard Model it is not \cite{marciano,amaldi1}. 
This was spectacularly confirmed in 1990 with the precise LEP1
measurements of the gauge couplings constants \cite{ellis,amaldi2}. 
This is the most compelling ``experimental'' indication for
supersymmetry and has lead to a flourish of activity on unification
and supersymmetry \cite{ohman,ellis,susyunif}. These studies focused
on the minimal supersymmetric Standard Model (MSSM) which is minimal
in particle content {\it and} \/couplings and conserves the discrete and
multiplicative symmetry R-parity\footnote{B: Baryon number, L: Lepton
number, S: Spin.}  \cite{mssm}
\beq 
R_p=(-)^{3B+L+2S}.
\eeq
We refer to this model as the $R_p$-MSSM, \ie the $R_p$ conserving MSSM\@.

If we require a supersymmetric Standard Model which is only minimal in
particle content the superpotential is modified to allow for additional 
R-parity violating (\rpv) interactions which are given in full below in 
Eq.\eq{superpot}. The superpotential includes terms which violate baryon 
number and separate terms violating lepton-number. In order to avoid 
rapid proton decay either baryon number or lepton number must be
conserved but not necessarily both. We refer to a model which violates
just one of these symmetries as an \rpv-MSSM, \cite{dreiner}. Symmetries 
which can achieve this are for example baryon parity and lepton parity
\cite{hallsuz,bp}
\beq
B_p=(-)^{3B+2S},\quad L_p=(-)^{L+2S}. 
\lab{discrete}
\eeq 
Thus, both the $R_p$-MSSM and the \rpv-MSSM require a discrete
symmetry beyond $G_{SM}$ and are theoretically equally well motivated
\cite{dreiner}.

\subsection{R-parity Violation and Grand Unification}
Given the intense study of unification in the $R_p$-MSSM it is the
purpose of this paper to study the gauge coupling unification in the
\rpv-MSSM\@. At first sight, it might seem unnatural to study
unification within the \rpv-MSSM, since \rpv\ is not obtained in the
simplest GUT models. In $SU(5)$ for example, the dimension-four R-parity 
violating interactions are given by the operator
\beq 
{\ol\psi_i} {\ol \psi_j} {\chi_k}, 
\lab{rpvop}
\eeq 
where $i,j,k=1,2,3$ are generation indices, and ${\ol\psi_i},\;
{\chi_k}$ are the ${\ol {\bf 5}}$, and $\bf 10$ representations of
$SU(5)$ respectively. The operator \eq{rpvop} contains {\it all} \/the
cubic terms of Eq.\eq{superpot}, \ie both the baryon- and lepton-number
violating interactions. This leads to unacceptably rapid proton decay
or unnaturally small couplings ($\sim 10^{-13}$) and thus must not be
present. In $SO(10)$ and in $E_6$ grand unification the dimension-four
R-parity violating interactions are directly prohibited by gauge invariance.

It seems R-parity violation and GUTs are incompatible. The reason is
that any R-parity violating symmetry which is consistent with the
bounds on proton decay, such as baryon parity and lepton parity in
Eq.\eq{discrete}, assigns quarks and leptons different quantum
numbers. But in GUTs quarks and leptons are in common multiplets and
thus must have the same non-$SU(5)$ quantum numbers. This
contradiction is resolved once the GUT symmetry is broken, \ie for
energy scales below $M_{GUT}$. Once the $SU(5)$ symmetry is broken,
R-parity violating terms can be generated which are consistent with
proton decay.

In general, we do not expect a GUT to be the final theory, it leaves
many of the same questions unanswered as in the Standard Model. For
example GUTs do not include gravity and therefore it should be an
effective theory embedded in a more fundamental one, such as
M-theory. This more fundamental theory will lead to a set of
non-renormalisable operators at the GUT scale such as \cite{giudice}
\beq 
\frac{k}{M_X}{\ol\psi_i} {\ol\psi_j}{\chi_k}\Sigma.
\lab{nonrenorm} 
\eeq 
This operator is suppressed by a mass scale $M_X\gsim M_{GUT}$. Here,
$\Sigma$ is a scalar field in the adjoint representation of $SU(5)$
and $k$ is a dimensionless coupling constant.  ${\ol\psi_i}$,
${\ol\psi_j}$, ${\chi_k}$ and $\Sigma$ can be combined to $SU(5)$
invariants in several ways. When $\Sigma$ receives a non-zero vacuum
expectation value $SU(5)$ is broken and the operators in
Eq.\eq{nonrenorm} can generate a subset of the \rpv\ interactions in
the superpotential \eq{superpot}, which are consistent with bounds on
proton decay \cite{giudice}. Models of this nature have been
constructed for the gauge groups $SU(5)$
\cite{hallsuz,giudice,tamvakis,vissani}, $SU(5)\times U(1)$
\cite{hallbrahm,giudice,tamvakis} and $SO(10)$ \cite{giudice}.

Below the $SU(5)$ breaking-scale the operators \eq{nonrenorm} are
effectively dimension-four operators. Their dimensionless coupling
constant $k<\!\Sigma\!>/M_X$ will run, \ie it will be renormalised and
it will contribute to the running of the other couplings in the
theory.  Thus even though at first sight GUTs and R-parity violation
seem inconsistent, this is not the case. Unless prohibited by a
special symmetry, we expect to have R-parity violation via
non-renormalisable operators in {\it any} \/GUT\@. At low-energy, this will
manifest itself in (effective) tri-linear R-parity violating
contributions to the superpotential. Above the GUT scale we will
have an $SU(5)$ symmetric theory with for example one unified gauge
coupling constant.

\subsection{Unification and Fermion Masses}
One particular aspect of unification we will focus on below is the GUT
prediction $m_b(M_U)=m_\tau(M_U)$ which has been very successful
\cite{btauguys,fixedpoint1,sakis1}. We shall study the effect of
R-parity violation on this prediction.  If there are no
non-renormalisable operators leading to effective fermion masses below
$M_{GUT}$, or if these operators are highly suppressed then we expect
the Yukawa unification to still hold in the presence of R-parity
violation. This can typically be achieved by a discrete symmetry but
should be incorporated in a general theory of fermion masses (or
Yukawa couplings). If the non-renormalisable terms have the
form
\beq
W=h_{ij}^u(\Sigma)\chi_i\chi_jh_u+h_{ij}^{e,d}(\Sigma)\chi_i{\ol\psi}_jh_d,
\eeq
the mass predictions are maintained. Here $h_u,\,h_d$ are the $SU(5)$
$5$ and $\ol 5$ Higgs superfields, respectively. $h_{ij}^u,\,h_{ij}^
{e,d}$ are general functions of the adjoint Higgs field. When $SU(5)$
is broken and $<\!\Sigma\!>\not=0$, the usual mass terms are generated.

In the MSSM, if one requires the Yukawa couplings to unify this greatly
reduces the allowed region of the (supersymmetric) parameters. In
particular one obtains a strict relation between the running top mass
$m_t(m_t)$ and the ratio of the vacuum expectation values (vevs) of
the two Higgs doublets, $\tan\beta$ \cite{ohman,marcelacarlos}.  Given
the observed top quark mass \cite{cdf} this results in a prediction
for $\tan\beta \sim1-3$ or $\tan\beta \sim 55$. Does this prediction
still hold when allowing for R-parity violation? In
Ref. \cite{marco}, by allowing only the bi-linear lepton number
violating terms, it is shown that bottom-tau Yukawa unification can
occur for any value of $\tan \beta$. The bi-linear term induces a
tau-sneutrino vev, which introduces an additional parameter into the
relation between $\lambda_t$ and $m_t$, as compared to the
MSSM\@. Bottom-tau Yukawa unification is then obtained by varying the
stau vev, and therefore $\lambda_t$ (and hence $\tan\beta$). Here, we
will focus on the effect of the tri-linear \rpv\ terms upon the
bottom-tau unification scenario.  The third generation \rpv-couplings
enter the evolution of $m_t,\,m_b$, and $m_\tau$ at one loop and can
thus have a large effect. Thus if we allow for \rpv\ we expect the
strict predictions of the MSSM to be modified. In
Section~\ref{sec:btau} we shall analyse this effect and show that
bottom-tau Yukawa unification becomes viable for any value of $\tan
\beta$, each one corresponding to a particular value of an \rpv\
coupling.

There has been much work to predict the fermion masses at the
weak scale from a simple symmetry structure at the unification scale
\cite{fixedpoint1,fermmassguys}. It is possible that the fermion mass
structure is determined by a broken symmetry \cite{massibross} where
only the top-quark Yukawa coupling is allowed by the symmetry at
tree-level. Its value is put in by hand and is presumably of order
one. The other couplings are then determined dynamically through the
symmetry breaking model. Given such a model, we would then still
require a prediction for the top-quark Yukawa coupling. An intriguing
possibility is that this Yukawa coupling is given by an infra-red
(quasi) fixed point \cite{ir}. The low-energy value then depends only
very weakly on the high-energy initial value; the exact opposite of a
fine-tuning problem.  In supersymmetric GUTs with bottom-tau
unification one typically requires large values of $\lam_t\sim1$ close
to the IR quasi fixed-point. This has been studied in detail in
Refs.\cite{ohman, marcelacarlos,fixedpoint1,fixedpoint2,topfix}.  We
investigate the effect of the \rpv-couplings on the fixed point in
Section~\ref{sec:fp}. Similar to the case of bottom-tau unification in
section~\ref{sec:btau} we find fixed-point structures for the top
Yukawa coupling for any value of $\tan \beta$, although the focussing
behaviour can be weakened depending upon the particular coupling
introduced.

\subsection{Present Status}

There have been several previous studies of the renormalisation group
equations (RGEs) of the \rpv-Yukawa couplings
\cite{rogervern,roybrah,sher,ralf,decarlos}, which have all been at the
one-loop level. The main point of this paper is that we present the
two-loop equations for the first time.\footnote{The two-loop equations
have been presented before in \cite{dreinerpois}. This work contained
a sign error in the RGEs as pointed out by the authors of
\cite{decarlos} and remained unpublished since one author left the
field with the computer program.} In \cite{sher,roybrah} the unitarity bounds
on the couplings were determined at one-loop. These are still the best
bounds on some of the baryon-number violating couplings. Below we
update these bounds using the two-loop renormalisation group equations
(RGEs). In \cite{rogervern} the complete one-loop RGEs for the
dimensionless couplings were first presented and the fixed point
structure was studied. We differ slightly in philosophy by also
considering the Yukawa unification scenario as discussed above and
considering the fixed point structure at two-loop in the RGEs. In
\cite{decarlos} the full one-loop RGEs including the soft breaking
terms were presented. These were used to study the bounds from flavour
changing neutral currents. We do not here consider the RGEs for the
soft breaking terms and this work is thus complimentary to ours.
Several models have also been constructed implementing one-loop
equations including the soft-terms \cite{ralf}. 
Since our results are mainly
model independent we do not comment on this work here.

The most important effect which enters at two-loop is that the running
of the gauge couplings now depends on the \rpv-couplings. One might
expect this effect to be small. But for higher generations the bounds
on the \rpv-couplings are weak and the couplings can be of order the
electromagnetic coupling ($e\approx 0.30$) or more. In addition, most
bounds are presented for scalar fermion masses of $100\gev$ and become
weaker for higher masses. At present the best 1$\sigma$ empirical
bounds for the highest generation couplings and for relevant scalar
fermion masses of $1\tev$ are \cite{dreiner,gautam}
\barr
\lam_{323}<0.6\,\quad\lam'_{333}<2.6\,\quad{\lam''}_{323}<0.43^*,
\lab{bounds}
\earr
where the asterisk indicates the bound for a $100\gev$ mass, and does
not have a simple analytic description of the scaling with mass.  At
$1.5\tev$ the bound on $\lam_{323}$ \cite{bgh} is almost identical to
the perturbative limit obtained below in Section~\ref{sec:un}.  The
bound on $\lam'_{333}$ \cite{gautam} at $1\tev$, is obtained by
scaling and as such is meaningless since perturbation theory breaks
down below $M_U$ for such large values. The appropriate bound is thus
the perturbative limit, which we obtain in Section~\ref{sec:un}.  A
mass-independent bound on $\lam''_{323}$ was found in ref.\cite{prob}
from requiring perturbativity up to the scale $M_U$.  However, we show
below that the bound from perturbative unification is dependent upon
$\tan \beta$, and we calculate this dependence.  We shall thus explore
all three couplings to the perturbative limit.

The outline of the paper is as follows. In Section 2 we present the
full two-loop renormalisation group equations for the gauge couplings
and the superpotential parameters. In Section 3 we present the
specific RGEs assuming there is only one \rpv-operator present at a
time. In Section 4 we outline the procedure for our numerical
analysis. In Section 5 we use the specific equations to study the
effects of R-parity violation on the unification of the gauge
couplings. We focus on the main predictions of unification: the
unification scale, the value of the gauge coupling at the unification
scale and the value of the strong coupling at $M_Z$. In Section 6 we
study the effects of the third generation \rpv-couplings on $b$-$\tau$
unification. In Section 7 we study the Landau poles and the fixed
points of the top- and the \rpv-Yukawa couplings. In Section 8 we
present our conclusions.

\nsect{Renormalisation Group Equations \label{sec:rges}}
The chiral superfields of the $R_p$-MSSM and the \rpv-MSSM have the 
following $G_{SM}=SU(3)_c\times SU(2)_L\times U(1)_Y$ quantum numbers
\barr
L:&(1,2,-\half),\quad {\bar E}:&(1,1,1),\qquad\, Q:\,(3,2,\frac{1}{6}),\quad
{\bar U}:\,(3,1,\frac{2}{3}),\nonr\\ {\bar D}:&(3,1,-\frac{1}{3}),\quad
H_1:&(1,2,-\half),\quad  H_2:\,(1,2,\half).
\lab{fields}
\earr
In the following we shall apply the work of Martin and Vaughn (MV)
\cite{mv} to the general \rpv-MSSM superpotential. For the generic
superpotential, $W$, we closely follow their notation
\beq
W=Y^{\phi_\rho \phi_\sigma \phi_\delta}\phi_\rho\phi_\sigma\phi_\delta / 6,
\eeq
where $\phi_{\rho,\sigma,\delta}$ denote any chiral superfield of the
model. The indices ${\rho,\sigma,\delta}$ run over all gauge and
flavour components. The \rpv-MSSM superpotential is then given by
\barr 
W&=& \eps_{ab} \left[ (\ye)_{ij} L_i^a
H_1^b {\bar E}_j + (\yd)_{ij} Q_i^{ax} H_1^b {\bar D}_{jx} +
(\yu)_{ij} Q_i^{ax} H_2^b {\bar U}_{jx} \right]\nonr \\ &&+
\eps_{ab}\left[ \frac{1}{2}(\lame{k})_{ij} L_i^a L_j^b{\bar E}_k +
(\lamd{k})_{ij} L_i^a Q_j^{xb} {\bar D}_{kx} \right]+
\frac{1}{2}\eps_{xyz} (\lamu{i})_{jk} {\bar U}_i^x{\bar
D}_j^y{\bar D}^z_k \nonr\\&& + \eps_{ab} \left[\mu H_1^a H_2^b
+\kap^i L_i^a H_2^b \right].  \lab{superpot} 
\earr 
We denote an $SU(3)$ colour index of the fundamental representation by
$x,y,z=1,2,3$. The $SU(2)_L$ fundamental representation indices are
denoted by $a,b,c=1,2$ and the generation indices by $i,j,k=1,2,3$.
We have introduced the twelve $3\times3$ matrices
\beq 
\ye,\quad \yd,\quad
\yu,\quad \lame{k},\quad \lamd{k},\quad \lamu{i}, 
\lab{matrices} 
\eeq
for all the Yukawa couplings. This implies the following conventions
in the MV notation
\barr 
Y^{L_i^a Q^{bx}_j \bar{D}_{ky}}
&=& Y^{L_i^a \bar{D}_{ky} Q^{bx}_j} = Y^{\bar{D}_{ky} L_i^a Q^{bx}_j}
= Y^{Q^{bx}_j L_i^a \bar{D}_{ky}} \nonum  \\ 
&=& Y^{Q^{bx}_j
\bar{D}_{ky} L_i^a} = Y^{\bar{D}_{ky} Q^{bx}_j L_i^a} =
(\lamd{k})_{ij} \eps_{ab} \delta^{y}_{x},\\ 
Y^{L_i^a L_j^b\bar{E}_k} &=& Y^{L_i^a \bar{E}_k L_j^b} = 
Y^{\bar{E}_k L_i^a L_j^b} =
(\lame{k})_{ij} \eps_{ab} = -(\lame{k})_{ji} \eps_{ab},
\\ 
Y^{{\bar U}_{ix}{\bar D}_{jy}{\bar D}_{kz}}&=&Y^{{\bar
D}_{jy}{\bar U}_{ix}{\bar D}_{kz}}= Y^{{\bar D}_{jy}{\bar D}_{kz}{\bar
U}_{ix}}= \eps_{x y z} \left({\bf\Lambda}_{U^i}\right)_{jk} =
-\eps_{x y z} \left({\bf\Lambda}_{U^i}\right)_{kj}, 
\earr 
We denote the $G_{SM}$ gauge couplings by $g_3$, $g_2$, $g_1$. In
Appendix A we have collected several useful group theoretical formulas
pertaining to $G_{SM}$ and the above field content. Here we mention
that for $U(1)_Y$ we use the normalisation as in GUTs and thus use
$g_1=\sqrt{3/5}\,g_Y$.  More details are given in the appendix. We
define our notation for the Yukawa couplings via the superpotential
including all \rpv\ terms.

We now in turn study the dimensionless couplings and then briefly also
discuss the mass terms $\mu,\kap_i$. We do not consider the
soft-breaking terms here.

\subsection{Gauge Couplings}
The renormalisation group equations for the gauge couplings are 
\beq
\frac{d}{dt}g_a = \frac{g_a^3}{16\pi^2} B_a^{(1)}
+\frac{g_a^3}{(16\pi^2)^2} \left[ \sum_{b=1}^3 B_{ab}^{(2)} g_b^2
-\sum_{x=u,d,e} \left(C_a^x \mbox{Tr}(\Y_x^\dagger \Y_x)+
A_a^x \sum_{i=1}^3\mbox{Tr}(\Lam_{x_i}^\dagger \Lam_{x_i})\right)
\right].
\lab{gaugerg}
\eeq
The coefficients $B_a,\,B_{ab},$ and $C_a^x$ have been given
previously \cite{bjorkjones} and for completeness we present them in
the appendix. The \rpv-effects on the running of the gauge couplings
appear only at two-loop and are new. We obtain
\beq
A_{a}^{u,d,e}=\left( \begin{array}{ccc}
12/5 & 14/5 & 9/5 \\
0 & 6 & 1 \\
3 & 4 & 0
\end{array}
\right).
\eeq
This completes the equations for the running of the gauge coupling constants
at two-loop.

\subsection{Yukawa Couplings}
In general the renormalisation group equations for the Yukawa
couplings are given by \cite{mv}
\beq
\frac{d}{dt} Y^{ijk}=Y^{ijp}\left[\frac{1}{16\pi^2}\gamma_p^{(1)k}
+\frac{1}{(16\pi^2)^2}\gamma_p^{(2)k}\right]+(k\lra i)+(k\lra j),
\lab{rgyukawa}
\eeq
and the one- and two-loop anomalous dimensions are
\barr
\gamma_i^{(1)j}&=& \half Y_{ipq}Y^{jpq}-2\delta_i^j\sum_ag_a^2C_a(i),\\
\gamma_i^{(2)j}&=&-\half Y_{imn}Y^{npq}Y_{pqr}Y^{mrj}+Y_{ipq}Y^{jpq}
\sum_a g_a^2[2C_a(p)-C_a(i)]\nonr \\&&+2\delta_i^j\sum_ag_a^2
\left[g_a^2C_a(i)
S_a(R)+2\sum_bg_b^2C_a(i)C_b(i)-3g_a^2C_a(i)C(G_a)\right]. \lab{gamtwo}
\earr
We have denoted by $C_a(f)$ the quadratic Casimir of the 
representation $f$ of the gauge group $G_a$. $C(G)$ is an invariant of
the adjoint representation of the gauge group $G$ and $S_a(R)$ is the
second invariant of the representation $R$ in the gauge group $G_a$. 
These quantities are defined in the appendix and their specific values 
are given there as well. We now first give the explicit version of
Eq.\eq{rgyukawa} for the matrices \eq{matrices} in terms of the
anomalous dimensions, and then we present the explicit forms for
$\gamma^{(1)f_j}_{f_i},$ and $\gamma^{(2)f_j}_{f_i}$.

\subsubsection{RG-Equations \label{sec:RGES}}
The RGEs for the Yukawa couplings (including full family dependence)
are  given by
\barr
\Dt (\ye)_{ij} &=& (\ye)_{ik}\ggam{E_j}{E_k}
+(\ye)_{ij}\ggam{H_1}{H_1}
-(\lame{j})_{ki}\ggam{H_1}{L_k}
+(\ye)_{kj}\ggam{L_i}{L_k},\label{eq:ye}\\
\Dt (\yd)_{ij} &=& (\yd)_{ik}\ggam{D_j}{D_k}
  +(\yd)_{ij}\ggam{H_1}{H_1}
  -(\lamd{j})_{ki}\ggam{H_1}{L_k}
  +(\yd)_{kj}\ggam{Q_i}{Q_k},\label{eq:yd}\\
\Dt (\yu)_{ij} &=& (\yu)_{ik}\ggam{U_j}{U_k}\label{yu}
+(\yu)_{ij}\ggam{H_2}{H_2}
+(\yu)_{kj}\ggam{Q_i}{Q_k}, 
\\
\Dt (\lame{k})_{ij} &=& (\lame{l})_{ij}\ggam{E_k}{E_l}
+(\lame{k})_{il}\ggam{L_j}{L_l}
+(\ye)_{ik}\ggam{L_j}{H_1}
-(\lame{k})_{jl}\ggam{L_i}{L_l}-(\ye)_{jk}\ggam{L_i}{H_1}, \label{eq:lame}\\
\Dt (\lamd{k})_{ij} &=& (\lamd{l})_{ij}\ggam{D_k}{D_l}\lab{lamd}
+(\lamd{k})_{il}\ggam{Q_j}{Q_l}
+(\lamd{k})_{lj}\ggam{L_i}{L_l}
-(\yd)_{jk}\ggam{L_i}{H_1},\\
\Dt (\lamu{i})_{jk} &=& (\lamu{i})_{jl}\ggam{D_k}{D_l} \lab{lamu}
+(\lamu{i})_{lk}\ggam{D_j}{D_l}
+(\lamu{l})_{jk}\ggam{U_i}{U_l}.
\earr
At two-loop the anomalous dimensions are given by
\beq
\ggam{f_i}{f_j}=\yukgam{f_i}{f_j}.
\eeq

\subsubsection{Anomalous Dimensions}
The one-loop anomalous dimensions are given by
\barr
\gamma^{(1)L_j}_{L_i} &=&\left(\ye \ye^\dagg \right)_{ji}
+(\lame{q}\lame{q}^\dagg)_{ji} +3 (\lamd{q}\lamd{q}^\dagg)_{ji}
-\delta_i^j(\frac{3}{10}g_1^2+\frac{3}{2}g_2^2), \lab{gamll1}\\
\gamma^{(1)E_j}_{E_i} &=& 2 \left(\ye^\dagg \ye \right)_{ji}
+ \mbox{Tr}(\lame{j}\lame{i}^\dagg) -\delta_i^j(\frac{6}{5}g_1^2),
\lab{gamee1}\\
\gamma^{(1)Q_j}_{Q_i} &=& \left(\yd \yd^\dagg \right)_{ji}
+ \left(\yu \yu^\dagg \right)_{ji}
+ (\lamd{q}^\dagg\lamd{q})_{ij}
-\delta_i^j(\frac{1}{30}g_1^2+\frac{3}{2}g_2^2+\frac{8}{3}g_3^2),\\
\gamma^{(1)D_j}_{D_i} &=& 2 \left(\yd^\dagg \yd \right)_{ij}
+2 \mbox{Tr}(\lamd{i}^\dagg\lamd{j})
+2 (\lamu{q}\lamu{q}^\dagg)_{ji}
-\delta_i^j(\frac{2}{15}g_1^2+\frac{8}{3}g_3^2)),\\
\gamma^{(1)U_j}_{U_i} &=& 2\left(\yu^\dagg \yu \right)_{ij}
+ \mbox{Tr}(\lamu{j}\lamu{i}^\dagg)
-\delta_i^j(\frac{8}{15}g_1^2+\frac{8}{3}g_3^2)),\\
\gamma^{(1)H_1}_{H_1} &=& \mbox{Tr}\left(3\yd\yd^\dagg+\ye\ye^\dagg \right)
-(\frac{3}{10}g_1^2+\frac{3}{2}g_2^2),\\
\gamma^{(1)H_2}_{H_2} &=& 3 \mbox{Tr}\left( \yu\yu^\dagg\right)
-(\frac{3}{10}g_1^2+\frac{3}{2}g_2^2),\\
\gamma^{(1)H_1}_{L_i} &=& {\gamma^{(1)L_i}_{H_1}}^* =-3 (\lamd{q}^*\yd)_{iq}
- (\lame{q}^*\ye)_{iq}.
\earr
Note that here, $H_{1,2},L,Q$ represent the fields $H_{1,2}^a$, $L^a$,
$Q^{xa}$ where $a$ is the index of the fundamental representation of
$SU(2)$ (i.e.\ no factors of $\eps_{ab}$ are factored in).  For
the two-loop anomalous dimensions we write
\beq
\gamma_{f_i}^{(2)f_j}=\left(\gamma_{f_i}^{(2)f_j}\right)_{yukawa}+
\left(\gamma_{f_i}^{(2)f_j}\right)_{g-y}+
\left(\gamma_{f_i}^{(2)f_j}\right)_{gauge}.
\eeq
These correspond respectively to the three terms of \eq{gamtwo}. These are
given explicitly below. The pure gauge two-loop anomalous dimensions are
given by
\barr
\left(\gamma^{(2){L_j}}_{L_i}\right)_{gauge} &=& \delta_i^j
(\frac{15}{4}g_2^4+\frac{207}{100}g_1^4+\frac{9}{10}g_2^2g_1^2),\\
\left(\gamma^{(2){E_j}}_{E_i}\right)_{gauge} &=& \delta_i^j\frac{234}{25}
g_1^4,\\
\left(\gamma^{(2){Q_j}}_{Q_i}\right)_{gauge}&=&\delta_i^j (
-\frac{8}{9}g_3^4+\frac{15}{4}g_2^4+\frac{199}{900}g_1^4 + 8g_3^2 g_2^2 +
\frac{8}{45} g_3^2g_1^2 + \frac{1}{10}g_2^2g_1^2),\\
\left(\gamma^{(2){D_j}}_{D_i}\right)_{gauge} &=&\delta_i^j (
-\frac{8}{9}g_3^4+\frac{202}{225}g_1^4 +\frac{32}{45} g_3^2g_1^2),\\
\left(\gamma^{(2){U_j}}_{U_i}\right)_{gauge} &=&\delta_i^j (
-\frac{8}{9}g_3^4+\frac{856}{225}g_1^4 +\frac{128}{45} g_3^2g_1^2),\\
\left(\gamma^{(2){H_1}}_{H_1}\right)_{gauge} &=&
\left(\gamma^{(2){H_2}}_{H_2}\right)_{gauge} =
\left(\gamma^{(2){L_j}}_{L_i}\right)_{gauge}, \\
\left(\gamma^{(2){H_1}}_{L_i}\right)_{gauge}
&=&\left(\gamma^{(2){L_i}}_{H_1}\right)_{gauge} = 0,
\earr
The mixed gauge-Yukawa two-loop anomalous dimensions are given by
\barr
\left(\gamma^{(2){L_j}}_{L_i}\right)_{g-y} &=& (16g_3^2-\frac{2}{5}g_1^2)
\left(\lamd{q}\lamdd{q} \right)_{ji} + \frac{6}{5} g_1^2 (\ye\ye^\dagg
+\lame{q}\lame{q}^\dagg)_{ji},
\\
\left(\gamma^{(2){E_j}}_{E_i}\right)_{g-y} &=&(6g_2^2-\frac{6}{5}g_1^2)
(\ye^\dagg\ye)_{ij} +(3g_2^2-\frac{3}{5}g_1^2) \mbox{Tr}(\lame{j}\lame{i}
^\dagg),
\\
\left(\gamma^{(2){Q_j}}_{Q_i}\right)_{g-y} &=& \frac{2}{5}g_1^2
[\left( \yd\yd^\dagg+2\yu\yu^\dagg\right)_{ji}
+\left(\lamd{q}^\dagg\lamd{q}\right)_{ij} ],
\\
\left(\gamma^{(2){D_j}}_{D_i}\right)_{g-y} &=&(\frac{16}{3}g_3^2+
\frac{16}{15}
g_1^2)\left(\lamu{q}\lamu{q}^\dagg \right)_{ji}\nonr \\
&&+(6g_2^2+\frac{2}{5}g_1^2)[\left(\yd\yd^\dagg\right)_{ji}+
\mbox{Tr}(\lamd{j}\lamd{i}^\dagg)],
\\
\left(\gamma^{(2){U_j}}_{U_i}\right)_{g-y} &=& (6g_2^2-\frac{2}{5}g_1^2)
\left(\yu^\dagg\yu\right)_{ij}+
(\frac{8}{3}g_3^2-\frac{4}{15}g_1^2)\mbox{Tr}(\lamu{j}\lamu{i}^\dagg),\\
\left(\gamma^{(2){H_1}}_{H_1}\right)_{g-y} &=& (16g_3^2-\frac{2}{5}g_1^2)
\mbox{Tr}(\yd\yd^\dagg) + \frac{6}{5}g_1^2 \mbox{Tr}(\ye\ye^\dagg), \\
\left(\gamma^{(2){H_2}}_{H_2}\right)_{g-y} &=& (16g_3^2+\frac{4}{5}g_1^2)
\mbox{Tr}\left(\yu\yu^\dagg\right),\\
\left(\gamma^{(2){H_1}}_{L_i}\right)_{g-y} &=&
\left(\gamma^{(2){L_i}}_{H_1}\right)_{g-y}^*=  
(\frac{2}{5}g_1^2-16g_3^2)
\left(\lamd{q}^*\yd\right)_{iq} -\frac{6}{5}g_1^2\left(\lame{q}^*\ye
\right)_{iq}.
\earr
The pure Yukawa two-loop anomalous dimensions are given by
\barr
-\left(\gamma^{(2){L_j}}_{L_i}\right)_{yukawa} &=& 2\lp\ye\yed\ye\yed\rp_{ji}+
\lp\yed\rp_{ki}\lp\ye\rp_{jl}\mbox{Tr}\lp\lamed{l}\lame{k}\rp \nonr \\
&+&
2\lp\lame{l}\lamed{k}\rp_{ji}\lp\yed\ye\rp_{lk}+
\lp\lame{l}\lamed{k}\rp_{ji} \mbox{Tr}\lp\lamed{l}\lame{k}\rp \nonr \\
&+&\lp\ye\yed\rp_{ji}\mbox{Tr}\lp\ye\yed+3\yd\ydd\rp  \\
&+&\lp\ye\rp_{jk}\lp 3\lamed{k}\lamd{p}\yds
+\lamed{k}\lame{p}\yes\rp_{ip} \nonr \\
&-&\lp\yes\rp_{ik}\lp 3\lame{k}\lamds{p}\yd
+\lame{k}\lames{p}\ye\rp_{jp} \nonr \\
&+&\lp \lamed{k}\ye\yed\lame{k} + 3 \lamed{k} \lamd{p} \lamdd{p} \lame{k} +
\lamed{k} \lame{p} \lamed{p} \lame{k} \rp_{ij}  \nonr \\
&+&6\lp\lamd{l}\lamdd{k}\rp_{ji}\left[\lp\ydd\yd\rp_{lk} + 
\mbox{Tr}\lp\lamdd{l}\lamd{k}\rp \
+\lp\lamu{q}\lamud{q}
\rp_{kl} \right]\nonr \\
&+&
3\lp \lamds{k} \yd \ydd \lamdt{k} + \lamds{k} \yu \yud \lamdt{k} \rp_{ij}+
3 \lp \lamd{k} \lamdd{p} \lamd{p} \lamdd{k} \rp_{ji},\nonr \\
-\left(\gamma^{(2){E_j}}_{E_i}\right)_{yukawa} &=&
2 \lp\yed\ye
\yed\ye 
+\yed\lame{l}\lame{l}^\dagg\ye 
+3\yed\lamd{l}\lamdd{l}\ye\rp_{ij}\nonr \\ 
&+&
2\left(\ye^\dagg\ye
\right)_{ij}\mbox{Tr}\left(\ye^\dagg\ye+3\yd\yd^\dagg\right)  \\
&-&2 \lp 3 \yed \lame{j} \lamds{m} \yd - \yed \lame{j} \lamed{m} \ye
\rp_{im} \nonr \\
&+& 2\mbox{Tr}\left[\lame{j}\lamed{i} \lp \ye\yed + \lame{l} \lamed{l} + 3
\lamd{l} \lamdd{l} \rp \right] \nonr \\
&-&2 \lp 3 \ydd \lamdt{m} \lamed{i} \ye - \yed \lame{m} \lamed{i} \ye \rp_{mj}
\nonr\\
-\left(\gamma^{(2){Q_j}}_{Q_i}\right)_{yukawa} &=&
2\lp\yd\ydd\yd\ydd\rp_{ji}
+\lp\yd\ydd\rp_{ji}\mbox{Tr}\lp\yed\ye+3\ydd\yd\rp \nonr \\
&+&2\lp\yu\yud\yu\yud\rp_{ji}+3\lp\yu\yud\rp_{ji}\mbox{Tr}\lp\yud\yu\rp
\nonr \\
&+&2\lp\lamdd{l}\lamd{m}\rp_{ij}\left[
\lp\ydd\yd\rp_{ml}+\lp\lamu{q}\lamud{q}\rp_{lm}+
\mbox{Tr}\lp\lamdd{m}\lamd{l}\rp\right] \nonr \\
&+& \lp\lamdd{m}\ye\yed\lamd{m}\rp_{ij}
+ 3\lp\lamdd{m}\lamd{q}\lamdd{q}\lamd{m}\rp_{ij}+
\lp\lamdd{m}\lame{q}\lamed{q}\lamd{m}\rp_{ij} \nonr \\
&+& \lp \yd \rp_{jl} \lp 3 \lamdd{l} \lamd{m} \yds +
\lamdd{l}\lame{m}\yes \rp_{im}
+2\lp\yd\lamud{q}\lamu{q}\ydd\rp_{ji} \nonr \\
&+&2\lp\yd\rp_{jm}\lp\ydd\rp_{li} \mbox{Tr}\lp\lamdd{m}\lamd{l}\rp  \\
&+&\lp\yds\rp_{il}
\lp 3\lamdt{l}\lamds{m}\yd 
+\lamdt{l}\lames{m}\ye\rp_{jm} 
+\lp\yu\rp_{jl}\lp\yud\rp_{ki}\mbox{Tr}\lp\lamud{l}\lamu{k}\rp \nonr \\
-\left(\gamma^{(2){D_j}}_{D_i}\right)_{yukawa} &=&
2\lp\ydd\yd\ydd\yd\rp_{ij}+2\lp\ydd\yu\yud\yd\rp_{ij}\nonr \\
&+&2\lp\ydd\yd\rp_{ij}\mbox{Tr}\lp\yed\ye+3\ydd\yd\rp
+ 2 \lp \ydt \lamdd{p} \lamd{p} \yds \rp_{ji} \nonr \\
&+& 2
\mbox{Tr}\lp\lamdt{j}\lamds{i}\lp\yd\ydd+\yu\yud\rp+\lamdd{i}\lamd{j}\lamdd{q} 
\lamd{q}\rp \nonr \\
&+&\mbox{Tr}\lp6\lamd{q}\lamdd{q}\lamd{j}\lamdd{i}+
2\lame{q}\lamed{q}\lamd{j} \lamdd{i}+2\ye\yed\lamd{j}\lamdd{i}\rp \nonr \\
&+& 4 \lp \lamu{m} \ydd \yd \lamud{m} \rp_{ji} + 4 \lp \lamud{m} \lamu{p}
\lamud{p} \lamu{m} \rp_{ij} \\
&-& 4 \lp \lamud{m} \rp_{ik} \lp \lamu{m} \rp_{jl}
\mbox{Tr} \lp \lamd{l} \lamdd{k} \rp 
+ \lp 6 \ydd \lamdt{j} \lamds{p} \yd \rp_{ip} \nonr \\ 
&+&2 \lp \ydd \lamdt{j} \lames{p} \ye
\rp_{ip}
+ \lp 6\ydt \lamdd{i} \lamd{p} \yds + 2  \ydt \lamdd{j} \lame{p}
\yes \rp_{jp} \nonr \\
&+& 2 \lp \lamud{k} \lamu{l} \rp_{ij} \left[ \mbox{Tr} \lp \lamud{l} \lamu{k}
\rp + 2 \lp \yud \yu \rp_{lk} \right]
\nonr \\
-\left(\gamma^{(2){U_j}}_{U_i}\right)_{yukawa} &=& 
2\lp \yud\yu\yud\yu
\rp_{ij}
+2\lp\yud\yd\ydd\yu\rp_{ij}+6\lp\yud\yu\rp_{ij}\mbox{Tr}\lp\yu\yud\rp \nonr \\
&+& 2 \lp \yud \lamdt{m} \lamds{m} \yu \rp_{ij}
+ 4 \lp  \lamud{i} \lamu{j} \rp_{lm} \mbox{Tr} \lp \lamdd{m} \lamd{l} \rp
\nonr \\ &+& 
4 \mbox{Tr} \lp \lamu{j} \lamud{i} \lamu{p} \lamud{p} \rp +
4 \mbox{Tr} \lp \lamud{i} \lamu{j} \ydd \yd \rp
\\
-\left(\gamma^{(2){H_1}}_{H_1}\right)_{yukawa} &=& 
\mbox{Tr}\lp 3\ye\yed\ye\yed
+9\ydd\yd\ydd\yd+3\yd\ydd\yu\yud\rp\nonr\\
&+&\mbox{Tr}\lp3\ye\yed\lamd{q}\lamdd{q}+\ye\yed\lame{q}\lamed{q}
+6\ydd\yd\lamud{q}\lamu{q} \right. \nonr \\
&+&\left.3\yd\ydd\lamdt{q}\lamds{q}\rp
+\lp\yed\ye\rp_{ji}\mbox{Tr}\lp\lamed{i}\lame{j}\rp \nonr \\
&+&6\lp\ydd\yd\rp_{ik}\mbox{Tr}\lp\lamdd{k} \lamd{i}\rp, \\
-\left(\gamma^{(2){H_2}}_{H_{2}}\right)_{yukawa} &=&
\mbox{Tr}\lp9\yu\yud\yu\yud+3\yu\yud\yd\ydd+3\yu\yud\lamdt{q}\lamds{q}\rp
\nonr \\
&+& 3\lp\yud\yu\rp_{jk}\mbox{Tr}\lp\lamud{k}\lamu{j}\rp,\lab{gam2hbhb}
\\
-\left(\gamma^{(2){H_1}}_{L_i}\right)_{yukawa} &=&
 \lp 3 \lamed{j} \lamd{p} \lamdd{p} \ye + \lamed{j} \lame{p} \lamed{p} \ye +
3\lamed{j} \ye \yed \ye \rp_{ij} \nonr \\
&+& \lp \lamed{k} \ye \rp_{il} \mbox{Tr}
\lp \lamed{l} \lame{k} \rp \nonr \\
&+&3 \lp \yes \yet \lamds{m} \yd - \yes \yet \lames{m} \ye
\rp_{im} - 9 \lp \lamds{k} \yd \ydd \yd \rp_{ik} \nonr \\ 
&+& 6 \lp \lamds{k} \yd \lamud{m} \lamu{m} \rp_{ik} - 6 \lp \lamds{k} \yd
\rp_{il} \mbox{Tr} \lp \lamd{l} \lamdd{k} \rp \nonr \\
&-& 3 \lp \lamds{j} \yu \yud \yd \rp_{ij} - 3 \lp \lamds{j} \lamdt{m}
\lamds{m} \yd \rp_{ij}
= -\left(\gamma^{(2){L_i}}_{H_1}\right)_{yukawa}^*
\earr
This completes the renormalisation group equations for the Yukawa
couplings at two-loop.  Before we discuss applications we briefly
consider the renormalisation of the bilinear terms.

\subsection{Bi-Linear Terms \label{sec:bi}}
Following the general equations given in MV the renormalisation group
equations for the bilinear terms now including all R-parity violating effects
are given by
\barr
\Dt\mu&=&\mu\left\{ \ggam{H_1}{H_1}+\ggam{H_2}{H_2}\right\}
+\kap^i\ggam{H_1}{L_i},\\
\Dt\kap^i&=&\kap^i\ggam{H_2}{H_2}+\kap^p\ggam{L_i}{L_p}
+\mu\ggam{L_i}{H_1}. \lab{kappa}
\earr
The anomalous dimensions at two-loop are given in the previous
subsection. We see that even if initially $\kap_i=0$ a non-zero
$\kap_i$ will in general be generated through the RGEs via a non-zero
$\mu$. As noted in MV the bi-linear terms do not appear in the
equations for the Yukawa couplings or the gauge couplings. They thus
do not directly affect unification and we ignore them for the rest of
this paper.

\subsection{Discussion \label{sec:dis}}
The two-loop renormalisation group equations for the Yukawa couplings
respect several symmetries. If at some scale for example $\lam''_{ijk
}=0$ for all $i,j,k$ then baryon parity, $B_p$, is conserved at this
scale. There are no \bpv-couplings in the theory and thus in
perturbation theory no \bpv-couplings are generated, \ie the RGEs
preserve $\lam''_{ijk}=0$ at all scales. Analogously, lepton parity,
once imposed, is also preserved by the RGEs.  If at some scale $\lam
_{ijk}=\lam''_{ijk}=0$ for all $i,j,k$ and only one lepton flavour is
violated, (\eg $\lam'_{3jk}\not=0$ in the mass basis of the charged
leptons) then this is also true for all scales. If the neutrino masses
are non-zero then this is no longer true~\cite{RPVneut}. The electron
mass matrix $\ye$ then contains off-diagonal entries which generate
off diagonal $\Gamma_{E_j}^{E_i},\Gamma_{L_j}^{L_i}$ via the RGEs. But
the effects will be very small and can thus be neglected in most
circumstances. However, if we assume only $\lam'_{111}\not=0$ at some
scale then through the quark CKM-mixing the other terms $\lam'_{1ij}$
will be generated.

Our results agree with MV for the MSSM Yukawa couplings. We also agree
with the one-loop $R_p$ violating results \cite{rogervern,prob,ralf}.

\nsect{Specific RGEs \label{sec:un}}
\subsection{Assumptions}
We now apply the two-loop RG-equations to the questions of unification. We
shall assume as a first approximation that the \rpv-couplings have a similar
hierarchy to the  SM Yukawa couplings and thus only consider one coupling at a
time. The third generation couplings have the weakest bounds \eq{bounds} and
can thus lead to the largest effects. We shall consider the three cases
\beq
{\bf LL\Ebar:}\,\;\lam_{323},\quad{\bf LQ\Dbar:}\,\;\lam'_{333},\quad{\bf \Ubar
\Dbar\Dbar:}\,\;\lam''_{323},
\lab{cases}
\eeq
defined in the current basis of the charged fermions.  We assume that in
each case the respective operator decouples from the other
\rpv-operators whose couplings we set to zero. Rigourously this is not
a consistent approach. However we can show that this is a very good 
approximation.

The coupling $(\Lam_{E^3})_{23}$ violates $L_\mu$ and can thus also
generate non-zero $(\Lam_{E^1})_{12}$ as well as $(\Lam_{D^k})_{2j}$.
We use $(\Lam_{E^3})_{23}$ as defined in a field basis in which $Y_E$
is diagonal, otherwise $L_{e,\tau}$ are not valid to protect the other
$(\Lam_{E^k})_{ij}$ from becoming non-zero.  The full RGEs are
directly determined from the equations in the previous section. The
largest relevant terms at one-loop order are
\barr
\frac{d}{dt}(\Lam_{E^1})_{12} &\approx& \frac{-1}{16\pi^2} 
(h_e h_\tau) (\Lam_{E^3})_{23}, \\
\frac{d}{dt}(\Lam_{D^3})_{23} &\approx& \frac{1}{16\pi^2} 
(h_b h_\tau) (\Lam_{E^3})_{23}. \lab{newcpl}
\earr
Here $h_e,\,h_\tau,\,h_b$ are the $e,\tau,$ and $b$ Yukawa couplings.
We do not know the entries in $Y_D$. Here we have assumed
$(Y_D)_{33}\approx h_b$. Then in both cases the newly generated couplings are 
strongly suppressed and we can safely drop them. As an additional
check, we have run the one loop RGEs in the above cases with 
$(\Lam_{E^3})_{23}$ non-zero at the unification scale. The initially
zero couplings are then generated at the $10^{-5}$ level even for
large values of $(\Lam_{E^3})_{23}|_{M_U}$.

The coupling $(\Lam_{D^3})_{33}$ violates $L_\tau$ and can in principle 
generate $(\Lam_{D^j})_{3i},\;(\Lam_{E^i})_{i3}\not=0$ at the one-loop level. 
The relevant terms in the RGEs are
\barr
\frac{d}{dt}(\Lam_{D^k})_{3j} &\sim& \frac{1}{16\pi^2}
(\Lam_{D^3})_{33}\left[2 \delta_{j3} 
(Y_D^\dagger Y_D)_{3k} + \delta_{k3} [(Y_DY_D^\dagger)_{3j}
+ (Y_UY_U^\dagger)_{3j}]
+3h_b (Y_D)_{jk}\right] , \\
\frac{d}{dt}(\Lam_{E^1})_{31} &\approx& \frac{3}{16\pi^2}
(h_e h_b ) (\Lam_{D^3})_{33}. \\
\frac{d}{dt}(\Lam_{E^2})_{32} &\approx& \frac{3}{16\pi^2}
(h_\mu h_b ) (\Lam_{D^3})_{33},
\earr
where $j,k$ are not equal to 3 simultaneously.
In order to estimate the contribution in the first RGE we would need
full knowledge of the matrices $Y_D$ and $Y_U$, which does not exist
to date.  If we use the symmetric texture ans\"atze of \cite{rrr} we
indeed find the off-diagonal elements of $Y_D$, $Y_D^\dagger Y_D$ and
$Y_U^\dagger Y_U$ highly suppressed. We thus feel justified in
neglecting the newly generated couplings. In the second and third case
we have made the same assumption as in \eq{newcpl} and we see that we
can safely ignore the newly generated couplings.

An initial non-zero $(\Lam_{U^3})_{23}$ can generate all the other 
baryon-number violating couplings. At one-loop and to leading order
these are $(\Lam_{U^i})_{jk}$, where $(ijk)\epsilon\{(312),(313),(123),(223)
\}$. The relevant terms in the RGEs are
\beq
\frac{d}{dt}(\Lam_{U^i})_{jk} \approx \frac{1}{8\pi^2}
(\Lam_{U^3})_{32}\left[
(Y_D^\dagger Y_D)_{31}\delta_{i3}\delta_{j2}\delta_{k1}+
(Y_D^\dagger Y_D)_{21}\delta_{i3}\delta_{j1}\delta_{k3}+
(Y_U^\dagger Y_U)_{3i}\delta_{j2}\delta_{k3}
\right].
\eeq
Under the previous assumptions this again leads to negligible new couplings. 
Thus in all three cases the decoupling is a good assumption. In line with this 
argument we assume the following form for the Higgs-Yukawa matrices
\beq
\ye=\mbox{diag}(0,0,\lam_\tau),\quad\yd=\mbox{diag}(0,0,\lam_b),\quad\yu=
\mbox{diag}(0,0,\lam_t).
\eeq
So we make the approximation that the current basis is equal to the
mass basis.

\subsection{Special Case Equations}
We now consider the specific case of the R-parity violation being
through one dominant operator as in \eq{cases}. Thus we set the
product of any two \rpv\ couplings to be zero. We note the connection
between the conventional notation and our matrix notation
\beq
(\lame{3})_{32}=\lam_{323}, \quad 
(\lamd{3})_{33}=\lam'_{333}, \quad
(\lamu{3})_{23}=\lam''_{323}.
\eeq
The gauge couplings have RGEs equivalent to those of the MSSM except for the
following two-loop \rpv\ contributions which we parameterise as  
\beq
\Delta \left(\Dt g_a\right)\equiv \Dt g_a^{\not\!\!R_p-MSSM} -\Dt
g_a^{R_p-MSSM}  
\eeq
where 
\barr
\Delta \left(\Dt g_3\right)  &=& -\frac{g_3^3}{(16 \pi^2)^2} \left[6 
{\lam''}_{323}^2 + 4 {\lam'}_{333}^2\right] \\ 
\Delta \left(\Dt g_2\right) &=& -\frac{g_2^3}{(16 \pi^2)^2} \left[ 6
{\lam'}_{333}^2 + 2 {\lam}_{323}^2 \right] \\
\Delta \left(\Dt g_1 \right) &=& -\frac{g_1^3}{(16 \pi^2)^2} \left[
\frac{24}{5} {\lam''}_{323}^2 + \frac{14}{5} {\lam'}_{333}^2 +
\frac{18}{5} \lam_{323}^2\right].
\earr
The RGEs of the Yukawa couplings are
\barr
\Dt \lam_{323} &=& \lam_{323} \left(\ggam{E_3}{E_3} + \ggam{L_3}{L_3} +  
\ggam{L_2}{L_2}\right) + \ltau \ggam{L_2}{H_1}  \lab{eq:splle}\\
\Dt {\lam'}_{333} &=& {\lam'}_{333} \left( \ggam{D_3}{D_3} + \ggam{Q_3}{Q_3} + 
\ggam{L_3}{L_3}\right) - \lambda_b \ggam{L_3}{H_1}  \lab{eq:splp}\\
\Dt {\lam''}_{323} &=& {\lam''}_{323} \left( \ggam{D_3}{D_3} + \ggam{D_2}{D_2} 
+ \ggam{U_3}{U_3}
\right)  \lab{eq:splpp}\\
\Dt \ltau &=& \ltau \left(\ggam{E_3}{E_3} +  \ggam{H_1}{H_1} +  \ggam{L_3}{L_3}
\right)+ \lam_{323} \ggam{H_1}{L_2}  \\
\Dt \lb &=& \lb \left( \ggam{D_3}{D_3} +  \ggam{H_1}{H_1} + 
\ggam{Q_3}{Q_3} \right) - \lam'_{333} \ggam{H_1}{L_3}
 \\
\Dt \lt &=& \lt \left(\ggam{U_3}{U_3} + \ggam{H_2}{H_2} + \ggam{Q_3}{Q_3}
\right). \lab{eq:splt}
\earr
The RGEs for the bi-linear terms are:
\barr
\Dt \mu &=& \mu \left( \ggam{H_1}{H_1} + \ggam{H_2}{H_2} \right) + \kap_2
\ggam{H_1}{L_2} + \kap_3 \ggam{H_1}{L_3}\\
\Dt \kap_1 &=& \kap_1 \left( \ggam{L_1}{L_1} +
\ggam{H_2}{H_2} \right) \\
\Dt \kap_2 &=& \mu \ggam{L_2}{H_1} + \kap_2 \left( \ggam{L_2}{L_2} +
\ggam{H_2}{H_2} \right) \\
\Dt \kap_3 &=& \mu \ggam{L_3}{H_1} + \kap_3 \left( \ggam{L_3}{L_3} +
\ggam{H_2}{H_2} \right). 
\earr
In the RGE for $\kap_1$ the term proportional to $\mu$ is dropped
because in our approximation $\ggam{L_1}{H_1}$ vanishes. The specific
two-loop anomalous dimensions are:
\barr
\ggam{E_3}{E_3} &=& \frac{1}{16 \pi^2} \left[ 2 \ltau^2 + 2 \lam_{323}^2 -
\frac{6}{5} 
g_1^2 \right]  \\&+& 
\frac{1}{(16 \pi^2)^2}
\left[ (\ltau^2 + \lam_{323}^2) ( -\frac{6}{5}g_1^2 + 6 g_2^2) - 
4 \ltau^4 - 8 \ltau^2 \lam_{323}^2\right. \nonum \\&-&\left. 6 \lb^2 
{\lam'}_{333}^2 
  - 6 \ltau^2 \lb^2 - 4 \lam_{323}^4 +
\frac{234}{25}g_1^4 \right] \nonum \\
\ggam{L_3}{L_3} &=& \frac{1}{16 \pi^2} \left[ \ltau^2 + \lam_{323}^2 + 3 
{\lam'}_{333}^2 -
\frac{3}{10} g_1^2 - \frac{3}{2} g_2^2\right] +\frac{1}{(16 \pi^2)^2}
\left[ \frac{15}{4} g_2^4 + \frac{207}{100} g_1^4 + \frac{9}{10} g_2^2 g_1^2 
\right. \\ &+& \left.
{\lam'}_{333}^2 (16 g_3^2 - \frac{2}{5} g_1^2) + \frac{6}{5} g_1^2 (\ltau^2 + 
\lam_{323}^2) -
3 \ltau^4 - 6 \ltau^2 \lam_{323}^2 - 3 \lam_{323}^4 - 3 \ltau^2 \lb^2 - 9
{\lam'}_{333}^4 \right. \nonum \\ &-&\left. 9 
{\lam'}_{333}^2 \lb^2 - 3 \lt^2 {\lam'}_{333}^2  \right]\nonum \\
\ggam{L_3}{H_1} &=& \ggam{H_1}{L_3}=-\frac{1}{16 \pi^2}3{\lam'}_{333} \lb   \\
&-&
\frac{1}{(16 \pi^2)^2}
\left[ {\lam'}_{333} \lb (16 g_3^2 - \frac{2}{5} g_1^2) + 3 \ltau^2 \lb
{\lam'}_{333} - 9 {\lam'}_{333}
\lb^3 - 9 {\lam'}_{333}^3 \lb - 3 {\lam'}_{333} \lt^2 \lb \right] \nonum \\
\ggam{L_2}{L_2} &=& \frac{1}{16 \pi^2} \left[ \lam_{323}^2 -
\frac{3}{10} g_1^2
 - \frac{3}{2} g_2^2 \right] \\ &+&\frac{1}{(16 \pi^2)^2}
\left[ \frac{15}{4} g_2^4 + \frac{207}{100} g_1^4 + \frac{9}{10} g_2^2 g_1^2 +
\frac{6}{5} g_1^2 \lam_{323}^2 - 3 \lam_{323}^4 - 3 \ltau^2
\lam_{323}^2\right]  \nonum \\
\ggam{L_2}{H_1} &=& \ggam{H_1}{L_2} = \frac{1}{16 \pi^2} \lam_{323}
\ltau+\frac {1}{(16 \pi^2)^2} \left[ \lam_{323} \ltau \frac{6}{5}
g_1^2 - 3 \lam_{323}^3 \ltau - 3 \ltau^3 \lam_{323}\right]\\
\ggam{D_3}{D_3} &=& \frac{1}{16 \pi^2} \left[ 2 \lb^2 + 2
{\lam''}_{323}^2 + 2 {\lam'}_{333}^2 - \frac{2}{15} g_1^2 -
\frac{8}{3} g_3^2\right] \\
&+&\frac{1}{(16 \pi^2)^2}
\left[-\frac{8}{9}g_3^4 + \frac{202}{225}g_1^4 + \frac{32}{45} g_3^2 g_1^2 +
({\lam'}_{333}^2 + \lb^2)(\frac{2}{5} g_1^2 + 6 g_2^2) + {\lam''}_{323}^2 ( 
\frac{16}{15} g_1^2 +
\frac{16}{3} g_3^2 ) \right. \nonum \\ &-&  16 \lb^2 {\lam'}_{333}^2 -8 \lb^4 
- 2 \lb^2 \lt^2 - 8 {\lam'}_{333}^4 -
2 {\lam'}_{333}^2 \lt^2 - 2 \lb^2 \ltau^2 - 8 {\lam''}_{323}^4 
\nonum \\ 
&-& \left. 4 {\lam''}_{323}^2 \lt^2 - 2 \ltau^2
{\lam'}_{333}^2
\right]\nonum \\
\ggam{Q_3}{Q_3} &=& \frac{1}{16 \pi^2} \left[ \lb^2 + \lt^2 + {\lam'}_{333}^2 -
\frac{1}{30} g_1^2 - \frac{3}{2} g_2^2 - \frac{8}{3} g_3^2\right] \\&+&
\frac{1}{(16 \pi^2)^2}
\left[-\frac{8}{9} g_3^4 + \frac{15}{4} g_2^4 + \frac{199}{900}g_1^4 + 8 g_3^2
g_2^2 + \frac{8}{45} g_3^2 g_1^2 + \frac{1}{10} g_2^2 g_1^2 + \frac{2}{5}
g_1^2 ( \lb^2 + 2 \lt^2 
\right.  \nonum \\  &+&\left. 
{\lam'}_{333}^2)
- 5 {\lam'}_{333}^4 - 10 {\lam'}_{333}^2 \lb^2 - \lb^2 \ltau^2
- 5 \lb^4 - 5 \lt^4 - 2 {\lam''}_{323}^2 \lt^2 - {\lam'}_{333}^2 \ltau^2 - 2 
\lb^2 {\lam''}_{323}^2\right]\nonum \\
\ggam{D_2}{D_2} &=& \frac{1}{16 \pi^2} \left[ 2 {\lam''}_{323}^2 - 
\frac{2}{15} g_1^2 - \frac{8}{3}g_3^2\right]\\&+& \frac{1}{(16 \pi^2)^2}
\left[ -\frac{8}{9} g_3^4 + \frac{202}{225} g_1^4 + \frac{32}{45} g_3^2 g_1^2
+ {\lam''}_{323}^2 ( \frac{16}{15} g_1^2 + \frac{16}{3} g_3^2) - 4
{\lam''}_{323}^2 \lb^2 \right. \nonum \\ &-&\left. 8
{\lam''}_{323}^4 - 4 {\lam''}_{323}^2 \lt^2\right]\nonum \\
\ggam{H_1}{H_1} &=& \frac{1}{16 \pi^2} \left[ 3 \lb^2 + \ltau^2 - \frac{3}{10}
g_1^2 - \frac{3}{2} g_2^2\right]\\
&+& \frac{1}{(16 \pi^2)^2}
\left[ \frac{15}{4} g_2^4 + \frac{207}{100} g_1^4 + \frac{9}{10} g_2^2 g_1^2 +
\lb^2 ( 16 g_3^2 - \frac{2}{5} g_1^2 ) + \frac{6}{5} g_1^2 \ltau^2 - 3 \ltau^2
\lam_{323}^2 - 3 \ltau^4  \right. \nonum \\&-& \left. 3 \ltau^2
{\lam'}_{333}^2
 - 9 {\lam'}_{333}^2 \lb^2 - 6 {\lam''}_{323}^2 \lb^2 - 9
\lb^4 - 3 \lb^2 \lt^2\right]\nonum\\
\ggam{H_2}{H_2} &=& \frac{1}{16 \pi^2} \left[  3 \lt^2 - \frac{3}{10} g_1^2 -
\frac{3}{2} g_2^2\right]\\ &+& \frac{1}{(16 \pi^2)^2}
\left[ \frac{15}{4} g_2^4 +\frac{207}{100} g_1^4 + \frac{9}{10} g_2^2 g_1^2 +
\lt^2 ( \frac{4}{5} g_1^2 + 16 g_3^2 ) \right. \nonum \\&-& \left. 6
{\lam''}_{323}^2 \lt^2 - 9 \lt^4 - 3 \lt^2 
\lb^2 - 3 {\lam'}_{333}^2 \lt^2 \right] \nonum \\
\ggam{U_3}{U_3} &=& \frac{1}{16 \pi^2} \left[ 2{\lam''}_{323}^2 + 2 \lt^2 -
\frac{8}{15} g_1^2 - \frac{8}{3} g_3^2 \right] \\
&+& \frac{1}{(16 \pi^2)^2} \left[ -\frac{8}{9} g_3^4 + \frac{856}{225} g_1^4 +
\frac{128}{45} g_3^2 g_1^2 + \lt^2 \left( 6 g_2^2 - \frac{2}{5}g_1^2 \right) +
{\lam''}_{323}^2 \left( -\frac{8}{15} g_1^2 +
\frac{16}{3} g_3^2 \right) \right. \nonum \\ &-& \left. 8 {\lam''}_{323}^4 -  
4 {\lam''}_{323}^2 \lb^2 - 2
\lt^2 {\lam'}_{333}^2 - 2 \lt^2 \lb^2 - 8 \lt^4
\right]
\earr
\begin{figure}
\begin{center}
\leavevmode   
\hbox{\epsfxsize=16.7cm
\epsffile{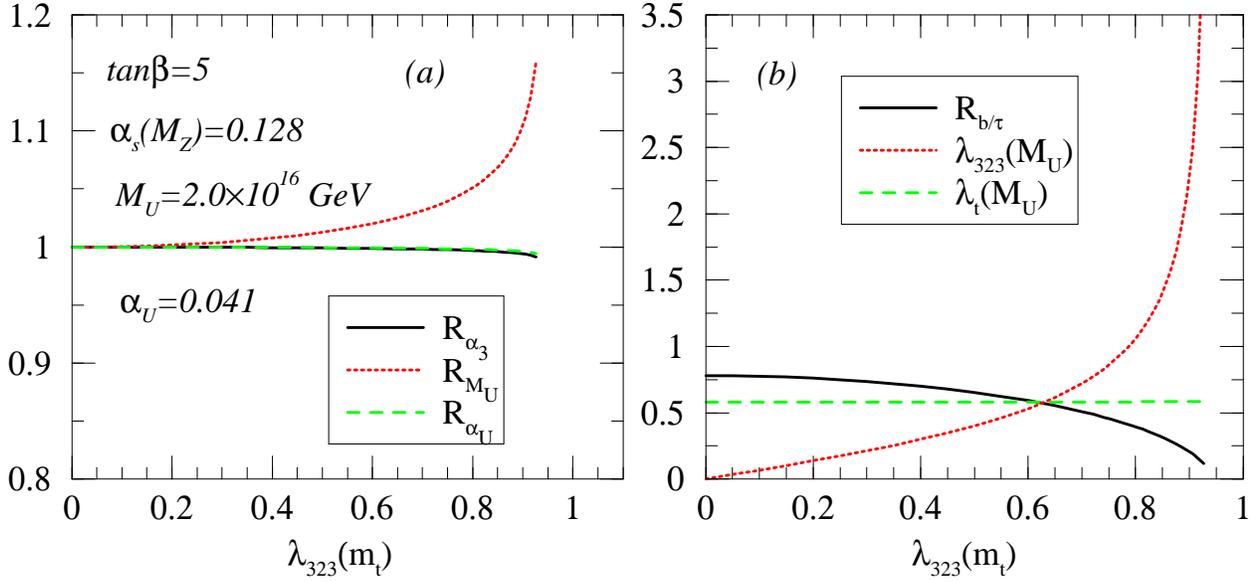}}
\end{center}
\caption{Effect of non-zero \rpv\ coupling $\lambda_{323}$ upon unification
predictions. The value of $\tan \beta$ input and the predictions for no \rpv\
are shown in the figures. (a) $R_i$ correspond to the ratio of the
prediction of $i$ in the $R_p$ case to the predictions in the \rpv\ case.
(b) $R_{b/\tau}=\lb(M_U)/\ltau(M_U)$.}
\label{fig:LE}
\end{figure}

\nsect{Outline of the Numerical Analysis of RGEs}
In order to determine the scale of unification we numerically solve
the renormalisation group equations.  In the process, we re-derive
unification predictions in the MSSM \cite{ohman,btauguys,marcelacarlos},
\cite{nirpaul}-\cite{sakis}. The main aim of this paper is to 
isolate the new effects due to the SUSY \rpv\ part of the theory. When
running the equations we must cross the mass thresholds of the
supersymmetric particles. In order to determine these we must also run
the two-loop RGEs of the soft supersymmetry breaking terms. The \rpv\
couplings contribute to these RGEs, but this contribution has not yet
been calculated at two-loop, although one-loop RGEs
exist~\cite{decarlos}.  In the future, when the full RGEs for the soft
terms have been calculated, it will be interesting to include the
\rpv\ effects on the spectrum and thus on the thresholds.  At this
stage, we make the approximation of using the $R_p$-MSSM RGEs for the
soft terms and the above \rpv\ RGEs for the dimensionless
couplings. We will also not consider (potentially large, but model
dependent) GUT threshold corrections \cite{polo,sakis}.

We add in turn one of the three \rpv-Yukawa couplings \eq{cases}.
We run the full set of equations including the two-loop
correction of the Yukawa couplings to the running of the gauge
couplings in Eq.~\eq{gaugerg}.  The boundary values of the running
$\overline{DR}$ gauge couplings $g_1(M_Z)$ and $g_2(M_Z)$ can be
determined in terms of the experimentally well known parameters, the
Fermi constant $G_F$, the Z-boson mass $M_Z$ and the electromagnetic
coupling $\alpha_{EM}^{-1}$ at $Q^2=0$. 
\barr
G_F&=& 1.16639\,10^{-5}\,GeV^{-2},\\
M_Z&=& 91.187\gev,\\
\alpha_{EM}^{-1}(Q^2=0)&=& 137.036.
\earr
For a given set of pole masses
$m_t=174\gev$, $m_b=4.9\gev$, $m_\tau=1.777\gev$ we define the
$\overline{DR}$ Yukawa couplings at $M_Z$.  Then we use 2-loop
Renormalisation Group equations to run up to the scale $M_U$, where
$g_1$ and $g_2$ meet\footnote{The gauge boson self energies $\Pi_{ZZ}
$, $\Pi_{WW}$ are not modified by the R-parity violating couplings up
to 1-loop level. We assume that the effect of those couplings in
vertex and box corrections to the running weak mixing angle is
negligible \cite{pierce,sakis}.}.

\begin{figure}
\begin{center}
\leavevmode   
\hbox{\epsfxsize=16.7cm
\epsffile{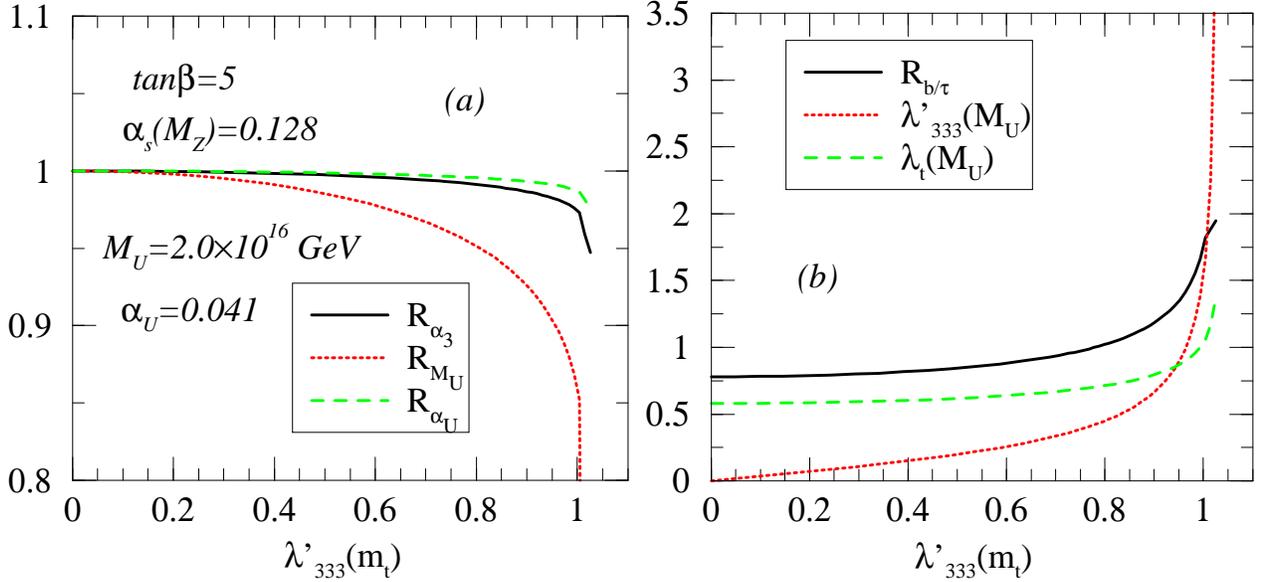}}
\end{center}
\caption{Effect of non-zero \rpv\ coupling $\lambda_{333}'$ upon unification
predictions. The value of $\tan \beta$ input and the predictions for no \rpv\
are shown in the figures. (a) $R_i$ correspond to the ratio of the
prediction of $i$ in the $R_p$ case to the predictions in the \rpv\ case.
(b) $R_{b/\tau}=\lb(M_U)/\ltau(M_U)$.}
\label{fig:LD}
\end{figure}

We have assumed universal boundary conditions at $M_U$ for the soft
breaking parameters $A_0$, $M_0$, $M_{1/2}$. The iteration procedure
used to determine our results is described in \cite{sakis}. As a
characteristic mean value of the soft breaking parameters at $M_U$ we
choose $A_0=M_{1/2}=M_0=300\gev$. We solve the RGEs for different
values of the \rpv-coupling at $m_t$, starting from zero.  The maximal
value we consider is where the running coupling reaches the
perturbative limit at the unification scale. For most of our
discussion we take this to be $\lam(M_U)<\sqrt{4\pi}\approx3.5$.
However, we also show some results where $\lam(M_U)<5.0$.

There has been a large interest in the literature
\cite{ohman,marcelacarlos,btauguys,fixedpoint1,sakis1,nirpaul2} in the
restrictions on the unification scenario from bottom-tau
unification. Requiring bottom-tau unification leads to a strict
relation between the running top quark mass and $\tan\beta$. For the
experimental value of $m_t$ \cite{cdf}, $\tan \beta$ is predicted to
be very close to 1.5 or around 55 \cite{figmarcela}. We are interested
in how the effects of \rpv\ can relax this strict relation and allow a
larger range of $\tan\beta$. As a model scenario, we consider
$\tan\beta=5$ which is well away from the solutions in the MSSM\@.  We
show that this feature is general and that solutions may be obtained
for any value of $\tan \beta$, provided one tunes the value of the
\rpv\ coupling.  The \rpv\ couplings change the fixed-point (and
quasi-fixed point) structure of the top-Yukawa coupling's
evolution. This is investigated in detail for low values of $\tan
\beta$.

\begin{figure}
\begin{center}
\leavevmode   
\hbox{\epsfxsize=16.7cm
\epsffile{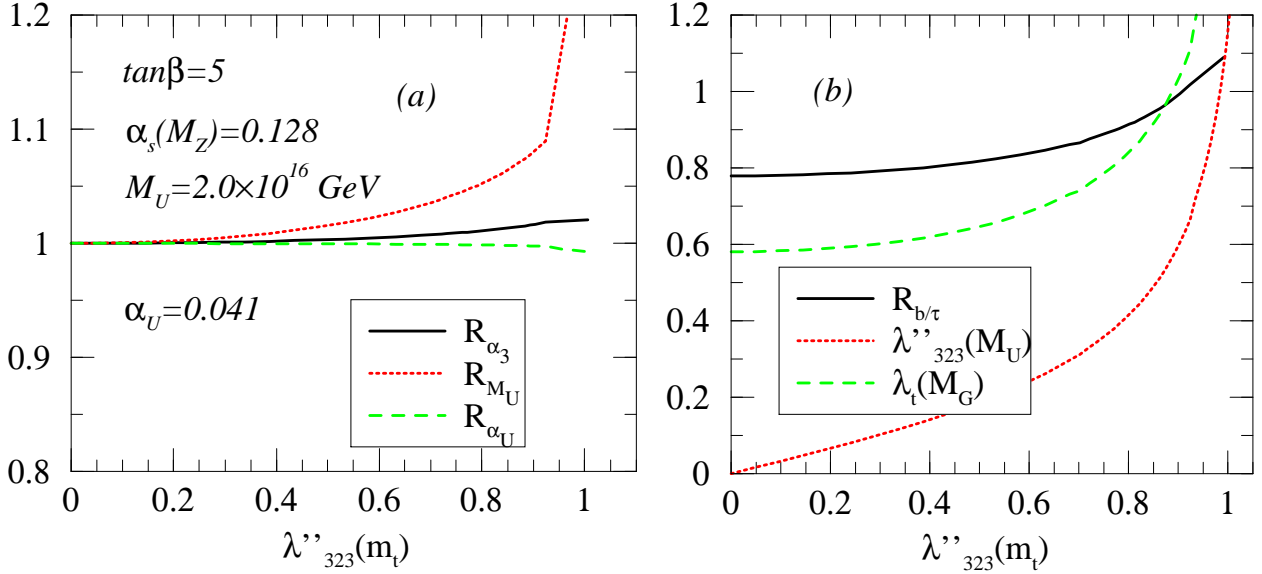}}
\end{center}
\caption{Effect of non-zero \rpv\ coupling $\lambda_{323}''$ upon unification
predictions. The value of $\tan \beta$ input and the predictions for the no
\rpv\ case
are shown in the figures. (a) $R_i$ correspond to the ratio of the
prediction of $i$ in the $R_p$ case to the predictions in the \rpv\ case.
(b) $R_{b/\tau}=\lb(M_U)/\ltau(M_U)$.}
\label{fig:LU}
\end{figure}

\nsect{Gauge Coupling Unification \label{sec:gauge}} 
For $\lambda_{\not R_p}=0$ and with the inputs and procedure defined above, 
we obtain  
\beq 
\alpha_s(M_Z)=0.128, \quad M_U=2.0\,10^{16}\gev,\quad \alpha_U=0.041. 
\lab{unifvalues} 
\eeq 
In order to discuss the effects of the non-zero \rpv\ Yukawa couplings we 
consider the unification parameters as functions of
$\lam_{\not R_p}$ evaluated at $m_t$. 
\beq 
\alpha_s(M_Z,\lam_{\not R_p}),\quad 
M_U(\lam_{\not R_p}),\quad\alpha_U(\lam_{\not R_p}). 
\eeq 
and define the ratios 
\barr 
R_{\alpha_3}(\lam_{\not R_p})&=&\frac{\alpha_3(M_Z,\lam_{\not 
R_p})}{\alpha_3(M_Z,0)},\nonr\\ 
R_{M_U}(\lam_{\not R_p})&=&\frac{M_U(\lam_{\not
R_p})}{M_U(0)}, 
\lab{ratios}\\ 
R_{\alpha_U}(\lam_{\not R_p})&=&\frac{\alpha_U(\lam_{\not R_p})}
{\alpha_U(0)}.\nonr 
\earr 
When $\lam_{\not R_p}=0$ and $\tan\beta=5$, the ratios
in \eq{ratios} are then all equal to 1 as can be seen on the left of
Figs.~\ref{fig:LE}a,~\ref{fig:LD}a, and~\ref{fig:LU}a, respectively.
Next we turn on the \rpv-couplings. In Fig.s~\ref{fig:LE}b,~\ref{fig:LD}b,
and~\ref{fig:LU}b we can read off the value of the \rpv-coupling at the 
unification scale as a function of the coupling at $m_t$. The plots stop 
once the perturbative limits are reached. For the present numerical
discussion we focus on $\tan\beta=5$. $\lam_{323}(M_U)$ reaches its
perturbative limit for a low scale value of $\lam_{323}(m_t)=0.93$.
It is worth pointing out that this is the same as the laboratory bound for
slepton masses at $1.5\tev$! Thus although the laboratory bounds on the
$LL\Ebar$ operators are generally considered to be very strict; for heavy
supersymmetric masses they are no stricter than the perturbative limit.
At this point $\lambda_{323}$ has run off Fig.~\ref{fig:LE}b but it should be
clear how it extrapolates. The perturbative limits for
the other couplings are given by
\barr
{\lam'}_{333}(m_t)&=& 1.06 \lab{eq:pld}\\
{\lam''}_{323}(m_t)&=& 1.07 \lab{eq:ple}
\earr
The first limit \eq{eq:pld} is equivalent to the empirical 2 $\sigma$
limit for $700\gev$ squark masses. The 1$\sigma$ empirical bound on
${\lam''}_{323}$ is 0.43 for $100\gev$ squark mass, and so the limit
\eq{eq:ple} will be restrictive for somewhat higher masses.  These
limits are $\tan \beta$ dependent, and we leave a full discussion to
section~\ref{sec:pert}.

In Fig.s~\ref{fig:LE}a,~\ref{fig:LD}a,~\ref{fig:LU}a we show how the ratios
\eq{ratios} change as we turn on
the \rpv-couplings. For $\lambda_{323}\not=0$, $\alpha_s(M_Z)$ and $\alpha_U$
are practically unchanged except very close to the perturbative limit. However,
$M_U$ is shifted upwards by up to $10(15)\%$, where the parenthesised value
corresponds to choosing the perturbative limit to be $\lambda<5.0$. 
For ${\lam'}_{333}$ the downward shift in $\alpha_s(M_Z)$ is typically
1-2\%. At the extreme perturbative limit $\lambda<5.0$ the maximum shift is
is a decrease of $5\%$ in $\alpha_s(M_Z)$ giving a value of 
$\alpha_s(M_Z)=0.122$. This corresponds to an
agreement with the data $\alpha_S(M_Z)=0.119\pm0.002$~\cite{pdg} to
1.5$\sigma$, without using GUT threshold corrections.
While the importance of this result is obvious, caution in its
interpretation is required because
the correction to $\alpha_S(M_Z)$ is sensitive to where one places the limit
of perturbative believability. For example, if one chooses $\lam(M_U)<3.5$, one
only obtains a 3$\%$ decrease, still with significantly better agreement with
the data than the $R_p$ conserving case.
$\alpha_U$ is decreased slightly at this point.
However, $M_U$ is decreased by up to $20\%$. This effect
is significantly beyond the effect due to the top quark Yukawa coupling.
For ${\lam''}_{323}\not =0$, $\alpha_U$ remains practically unchanged.
$\alpha_s$ now has an overall increase of up to about $3\%$ at the
perturbative limit
corresponding to a value of $\alpha_s(M_Z)=0.131$ in disagreement with the
experimental value. $M_U$ is raised by up to $20\%$.

Thus we find $\alpha_U$ essentially unchanged by \rpv-effects. $M_U$
can change either way by up to $20\%$. If we compare this with other
effects considered in Ref. \cite{nirpaul} we find it of the same order
as the uncertainty due to the top quark Yukawa coupling or the effects
of possible non-renormalisable operators at beyond the GUT scale. The
effect is much smaller than that due to GUT-scale threshold
corrections or weak-scale supersymmetric threshold corrections. It is
thus much too small an effect to accommodate string unification. The
strong coupling can also change either way by up to $5\%$. A decrease
is favoured by the data and is welcome in supersymmetric unification.
The effect of the \rpv-couplings on $\alpha_s(M_Z)$ is of the same
order as the effects due to the top-quark Yukawa coupling, GUT-scale
threshold effects and high-scale non-renormalisable operators
\cite{nirpaul}.

\nsect{b-$\tau$ Unification \label{sec:btau}}
\begin{figure}
\begin{center}
\leavevmode   
\hbox{\epsfxsize=14.7cm
\epsffile{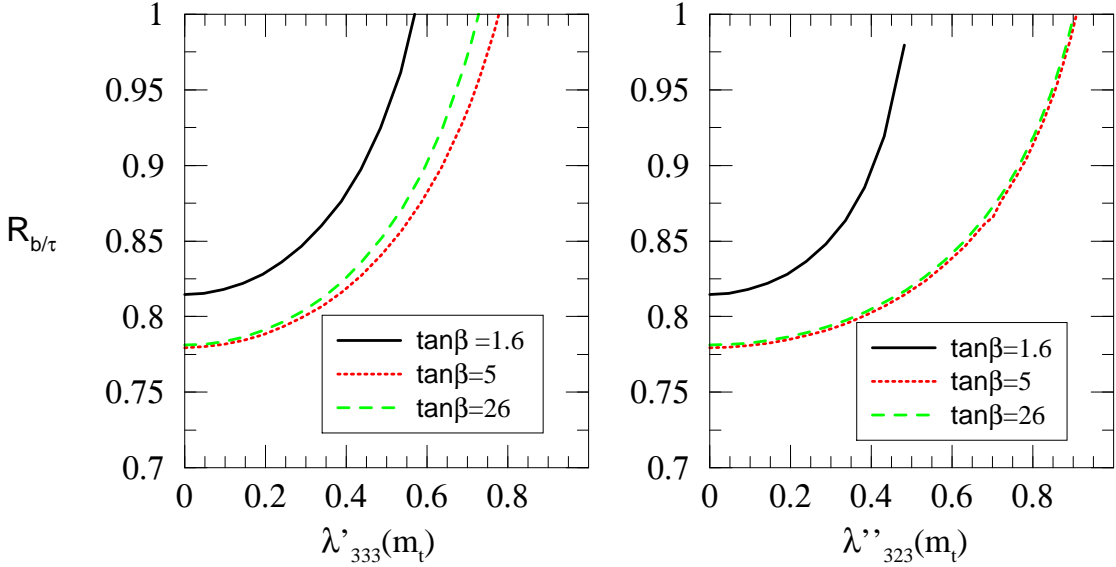}}
\end{center}
\caption{Bottom-tau Yukawa unification including the R-parity violating
couplings (a) $\lambda_{333}'$, (b) $\lam''_{323}$ for various $\tan
\beta=1.6, 5, 26$.
The plots show $R_{b/\tau} = \lambda_b(M_U) / \lambda_\tau(M_U)$ as a function
of the \rpv\ couplings evaluated at $m_t$. In (b), the $\tan \beta=1.6$ curve
stops as $\lambda_t(M_U)$ reaches its perturbative limit.}
\label{fig:btau}
\end{figure} 
In order to study the unification of the bottom and $\tau$ Yukawa couplings
$\lam_b,\,\lam_\tau$ at $M_U$ we define the ratio
\beq
R_{b/\tau}(M_U)=\frac{\lam_b(M_U,\lam_{\not R_p})}{\lam_\tau(M_U,\lam_{\not
R_p})}.
\eeq
For $\lam_{\not R_p}=0$, $\tan\beta=5$ we have 
\beq
R_{b/\tau}(M_U)=0.78.
\eeq
Thus including the top-quark effects but before turning on the \rpv-coupling
we are well away from the bottom-tau unification solution $R_{b/\tau}(M_U)=1$.
Recall that the uncertainties due to the bottom quark mass are small for small
$\tan\beta$. Now we consider the corrections due to the \rpv-couplings. The
one-loop RGE for $R_{b/\tau}(t)$ is given by
\beq
16 \pi^2 \frac{dR_{b/\tau}(t)}{dt} = R_{b/\tau}(t) \left[ \lam_t^2 + 3
\lam_b^2 - 3\lam_\tau^2
- 4\lam_{323}^2 + 3{\lam'}_{323}^2 + 2\lam^{''2}_{323} + \frac{4}{3} g_1^2 -
\frac{16}{3} g_3^2 \right]. \lab{Rbtau}
\eeq
The leading dependence of $R_{b/\tau}$
on $\lam_{323}$ has a negative sign and as we see in the two-loop result shown
in Fig.~\ref{fig:LE}b $R_{b/\tau}$ drops significantly. 
Near the perturbative limit it
drops by a factor of 2.5 and \rpv\  becomes a dominant effect on the
evolution of $R_{b/\tau}$. 
This is important for the range of $\tan\beta$ which
leads to bottom-tau unification. In the MSSM $R_{b/\tau}$ is too large for
$\tan\beta\lsim1.5$ or $\,\,\gsim55$ \cite{figmarcela}. Including a non-zero
operator $\lam_{323}$ strongly reduces $R_{b/\tau}$ and thus can lead to
bottom-tau unification in this previous regime.

For ${\lam''}_{323}\not=0$ or ${\lam'}_{333}\not=0$ there is an
additional {\it positive} \/contribution in the evolution of $R_{b/
\tau}(t)$. The full two-loop result shows a clear rise in $R_{b/
\tau}(t)$ as a function of ${\lam''}_{323}$ in Fig.~\ref{fig:LU}b. 
The maximum increase at the perturbative limit is by $60\%$. For $
{\lam''}_{323}(m_t)=0.9$ bottom-tau unification is restored! This is
quite remarkable. Even though \rpv-couplings are usually expected to
lead to only small effects they can have a significant impact on our
understanding of Yukawa-unification.  Recall, that grand unification
{\it is} \/possible in \rpv-theories as discussed extensively in the
introduction.  From Eq.~\eq{Rbtau} it should be clear that for example
for ${\lam'}_{333}$ we get an increase in $R_{b/\tau}(M_U)$ as well
leading to further bottom-tau unification points.  From
Fig.~\ref{fig:LD}, we see that for ${\lam'}_{333}(m_t) = 0.75$, we
achieve $R_{b/\tau}=1$ for $\tan \beta=5$.  To investigate further, we
plot $R_{b/\tau}$ as a function of ${\lam'}_{333}(m_t)$ for several
other values of $\tan \beta$ in Fig.~\ref{fig:btau}a. The figure
illustrates that any value of $\tan \beta=2-26$ achieves bottom-tau
Yukawa unification, provided ${\lam'}_{333}(m_t)$ is chosen
correctly. Fig.~\ref{fig:btau}b shows the equivalent plot for
${\lam''}_{323} \not =0$ with the same conclusions.

\nsect{Landau Poles and Fixed Points of Yukawa Couplings \label{sec:fp}}
True fixed points and quasi fixed points were first considered in
\cite{quasi}. Since then, many applications have been discussed 
in the literature (see ref.\cite{recentfp} and references therein),
including those in the $R_p$ conserving MSSM\@.  In the MSSM and at
low $\tan \beta$, it is well known that if one neglects two-loop
corrections as well as those from couplings smaller than $g_3$, the
RGE for $\lt$ has an infra-red stable fixed point \cite{ir}
corresponding to $\lt / g_3 = \sqrt{7/18}$. This is too low to
accommodate $m_t=175\pm 5\gev$.  To be phenomenologically viable,
$\lt$ must be out of the domain of attraction of the true
fixed-point. In practice, this implies that $\lt(m_t)$ must be nearer
to its quasi-fixed point (QFP) limit of 1.1, where $\lt(M_U)$ is large
(formally it diverges). Using $m_t(m_t)=165 \pm5\gev$ as an input
means that one can derive $\tan\beta$ in this scenario through the
relation
\beq
\sin \beta = \frac{\sqrt{2} m_t(m_t)}{v \lt(m_t)} \label{eq:tan},
\eeq
implying $\tan \beta = 1.7\pm 0.2$ at the QFP~\cite{haberespin}.  This
scenario is very attractive~\cite{fixedpoint2,recentfp} because the
values of many parameters in the infra-red regime are insensitive to
their input values at $M_U$, giving higher predictivity and a tightly
constrained phenomenology. The quasi-fixed $R_p$ conserving MSSM is
presently severely constrained~\cite{haberespin,recentfp}.  A large
\rpv\ coupling changes the running significantly, and so in this
section, we examine the running of the couplings as a function of
renormalisation scale.  It is convenient to split this analysis into
three cases to focus upon: (1) infra-red stable fixed points of
approximated RGEs, (2) a small amount of \rpv\ near the QFP of $\lt$
and (3) the case of two large Yukawa couplings. We then examine the
constraints on the \rpv\ couplings defined at $m_t$ coming from the
requirement of perturbativity up to $M_U$.

\subsection{Fixed Points}
First, we analyse the fixed points in the equations including the couplings
$\lam'_{333}, \lam''_{323}$ one at a time. 
Neglecting $\lb, \ltau$ (i.e.\ looking at the low $\tan \beta$ limit) and
two-loop terms, we reformulate Eqs.
\eq{eq:splt}, \eq{eq:splle}-\eq{eq:splpp} as
\barr
\frac{d \ln X_t}{d \ln g_3^2} &=& \frac{7}{9} - \frac{1}{3} X' - 2 X_t -
\frac{2}{3} X'' + \frac{\alpha_2}{\alpha_3} + \frac{13}{45} \frac{\alpha_1}
{\alpha_3}\\
\frac{d \ln X}{d \ln g_3^2} &=& -1 + \frac{3}{5}\frac{\alpha_1}{\alpha_3} +
\frac{\alpha_2}{\alpha_3} - \frac{4}{3}X \\
\frac{d \ln X'}{d \ln g_3^2} &=& \frac{7}{9} - \frac{1}{3} X_t - 2 X' +
\frac{\alpha_2}{\alpha_3} + \frac{7}{45} \frac{\alpha_1}{\alpha_3}\\
\frac{d \ln X''}{d \ln g_3^2} &=& \frac{5}{3} - 2 X'' - 
\frac{2}{3} X_t + \frac{4}{5} \frac{\alpha_1}{\alpha_3},
\earr
where $X_t \equiv \lambda_t^2 / g_3^2$, $X\equiv{\lam}_{323}^2/
g_3^2$, $X' \equiv{\lam'}_{333}^2 / g_3^2$ and $X'' \equiv{\lam
''}_{323}^2 / g_3^2$.  The stability of Yukawa couplings in
supersymmetric theories has been considered in refs.~\cite{stability}.
Considering only one non-zero \rpv\ coupling at a time, we have two
infra-red stable fixed points in the limit that we ignore the
electroweak gauge couplings. The first is
\beq
X=X'=0: \qquad X_t=\frac{1}{8}, \qquad X''=\frac{19}{24} 
\lab{eq:fp1}
\eeq
and the second is
\beq
X=X''=0: \qquad X_t = X' = \frac{1}{3}. 
\lab{eq:fp2}
\eeq
We have ignored $\lam_{323}$ in this discussion because it does not
exhibit fixed point behaviour itself due to the lack of
renormalisation from the QCD interactions. The values of $X_t$ in
Eqs. \eq{eq:fp1},\eq{eq:fp2} are even lower than the $R_p$-conserving
MSSM value $X_t=7/18$. They are experimentally excluded
\cite{recentfp} by the lower bound on $\lam_t(m_t)$ coming from the
requirement of $m_t(m_t)=165\gev$.

\subsection{Large $\lt(M_U)$, Small \rpv\ Coupling}
We now discuss the case where $\lt(M_U)>1$ and the \rpv-couplings are
in turn very small. The behaviour of the $R_p$
conserving superpotential parameters will be similar to their
behaviour in the $R_p$ conserving MSSM\@. In this case, we can solve the
one-loop RGEs for the \rpv\ Yukawa couplings analytically:
\barr
{\lam}_{323}(\mu) &=& {\lam}_{323}(M_U)
\delta_2^{-3} \delta_1^{-3/11}, \label{eq:qsol1}\\
{\lam'}_{333}(\mu) &=& {\lam'}_{333}(M_U)
\delta_3^{40/27} \delta_2^{-5/2} \delta_1^{-29/594} \delta_t^{1/6},\\
{\lam''}_{323}(\mu)&=& {\lam''}_{323}(M_U)
\delta_3^{56/27} \delta_2 \delta_1^{-95/297} \delta_t^{1/3}, \label{eq:qsol2}
\earr
where $\delta_{1,2,3} \equiv g_{1,2,3}(\mu) / g(M_U)$ and $\delta_t
\equiv \lt(\mu) / \lt(M_U)$. We have neglected contributions from
$\ltau, \lb$ in Eqs.~\eq{qsol1}-\eq{qsol2} and so they are valid only
at low $\tan \beta<15$. The one-loop analytic solutions for
$\delta_{1,2,3}$ are equivalent to the $R_p$ conserving MSSM ones
\cite{recentfp}.  The solution for $\delta_t$ is also equivalent to
its $R_p$ counterpart, which has been solved analytically at one loop
order including $g_3,g_2$~\cite{carab}.  Eqs.~\eq{qsol1}-\eq{qsol2}
predict a constant ratio of $\lam_{\not R_p}(M_U)/\lam_{\not R_p}
(m_t)$. This corresponds to a straight line for the $\lam_{\not R_p}
(M_U)$ curves in
Figs.~\ref{fig:LE}b,\,\ref{fig:LD}b,\,\ref{fig:LU}b. For $\lam_{\not
R_p}(m_t) \lsim 0.3$ this is a good approximation, after that the
two-loop effects become important. We also obtain analytic solutions
at low $\tan \beta$ for the bi-linear \rpv\ terms when the \rpv\
Yukawa couplings are small:
\beq
\kap_i (\mu) = \kap_i(M_U) \delta_1^{-5/198} \delta_2^{-3/2}
\delta_3^{-8/9} \delta_t^{1/2}.
\eeq
\subsection{Two large Yukawa couplings}
The one-loop RGEs including $\lt$, $\lb$, $g_{2,3}$ have been solved in the
MSSM~\cite{solns}. These RGEs have the same form in the case of a non-zero
$\lam'_{333}$, when ignoring all other Yukawa couplings except $\lt$.
The analytic solutions to this system are therefore contained in
ref.~\cite{solns} and are in terms of hypergeometric functions.
\begin{figure}
\begin{center}
\leavevmode   
\hbox{\epsfxsize=16cm
\epsffile{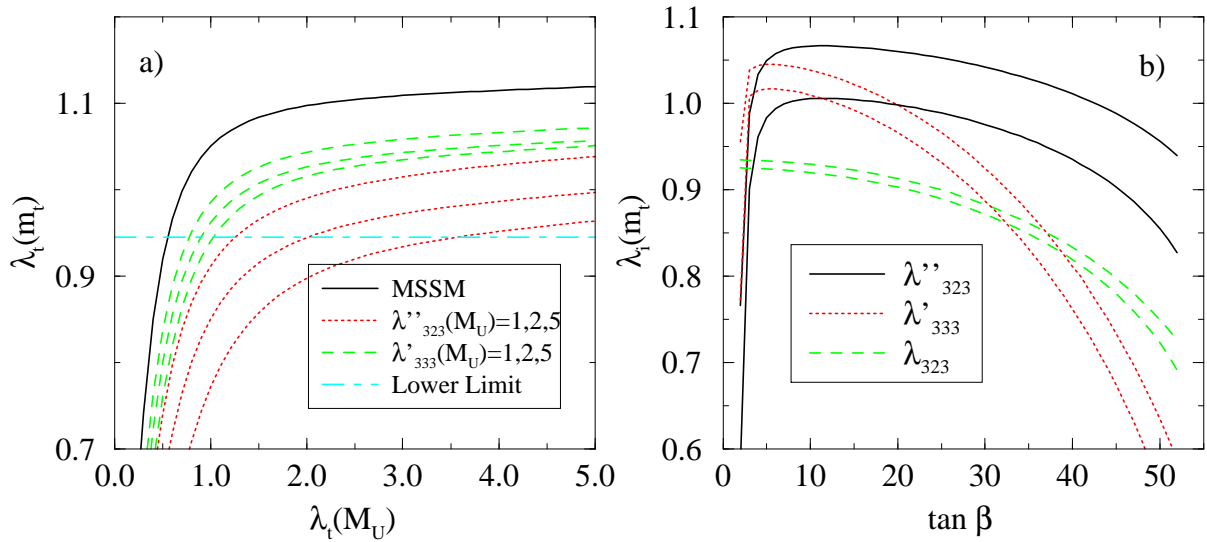}}
\end{center}
\caption{(a) Effect of large R-parity violating couplings upon the quasi-fixed
point of $\lambda_t$. The \rpv\ couplings increase toward the bottom of the
plot. The lower limit of $\lambda_t(m_t)$ is defined as the minimum value
required to obtain $m_t(m_t)=165\gev$. (b) Perturbative limits of large 
R-parity violating couplings as a function of $\tan \beta$. The upper curves 
correspond to the two-loop order calculation and the lower ones to a 
one-loop calculation.}
\label{fig:fixed}
\end{figure} 

We present here numerical solutions to the {\it two}-loop equations,
which can also be applied to the cases of large ${\lam}_{323}$, or
${\lam''}_{323}$ with large $\lt(m_t)$. Using $\alpha_U=0.041$, $M_U=2
\times 10^{16}\gev$ and again switching off $\lb$ and $\ltau$ (which
is only valid for low $\tan \beta\lsim 15$), we calculate how $\lt(m_t
)$ is related to its input value at $M_U$, \ie we study the
quasi-fixed point structure. SUSY threshold corrections are not
included.

In Fig.~\ref{fig:fixed}a, the top solid line shows the quasi-fixed
point structure of $\lt$ in the MSSM\@. For $\lt(M_U)>2.0$ it is almost
flat, i.e. $\lt(m_t)=1.1$ becomes insensitive to $\lt(M_U)$. The
horizontal dashed line corresponds to the minimum $\lt(m_t)$ required
to produce a top quark mass which agrees with the data and for which
$\sin\beta\leq1$. It changes by $\pm 0.03$ within the empirical errors
on $m_t$.  The $\lam_{323}\not = 0$ curves are not plotted because
they coincide with that of the MSSM\@.  This can be understood from
Eq.~\eq{eq:splt} as the coupling does not directly appear in the
running of $\lt$. This holds at two-loop due to our assumption of only
one dominant \rpv\ coupling at a time. 

When ${\lam'}_{333}$ is switched on, the quasi-fixed point behaviour
of $\lt$ persists but its QFP value can be decreased as far as $\lt
(m_t)= 1.03$ which corresponds to $\tan\beta=2.1\pm^{0.9}_{0.1}$. Any
decrease in $\lt(m_t)$ can increase $\tan \beta$ as extracted from
Eq.~\eq{tan}, modifying the QFP predictions of superpartner masses,
for example. In particular, higher $\tan \beta$ can allow for higher
masses of the lightest CP-even Higgs, relaxing a severe constraint in
the $R_p$ conserving QFP scenario~\cite{haberespin}.  For large ${
\lam''}_{323}$, the quasi-fixed behaviour of $\lt$ is somewhat
weakened as the non-zero gradient of the curves suggests, but $\lt
(m_t)=1$ is possible as the QFP prediction\footnote{This lowering of
$\lt(m_t)$ was first pointed out by Brahmachari and Roy in
ref.~\protect\cite{prob}}, corresponding to $\tan\beta=3.0\pm^{1.5}_
{0.7}$.

\begin{figure}
\begin{center}
\leavevmode   
\hbox{\epsfxsize=16.7cm
\epsffile{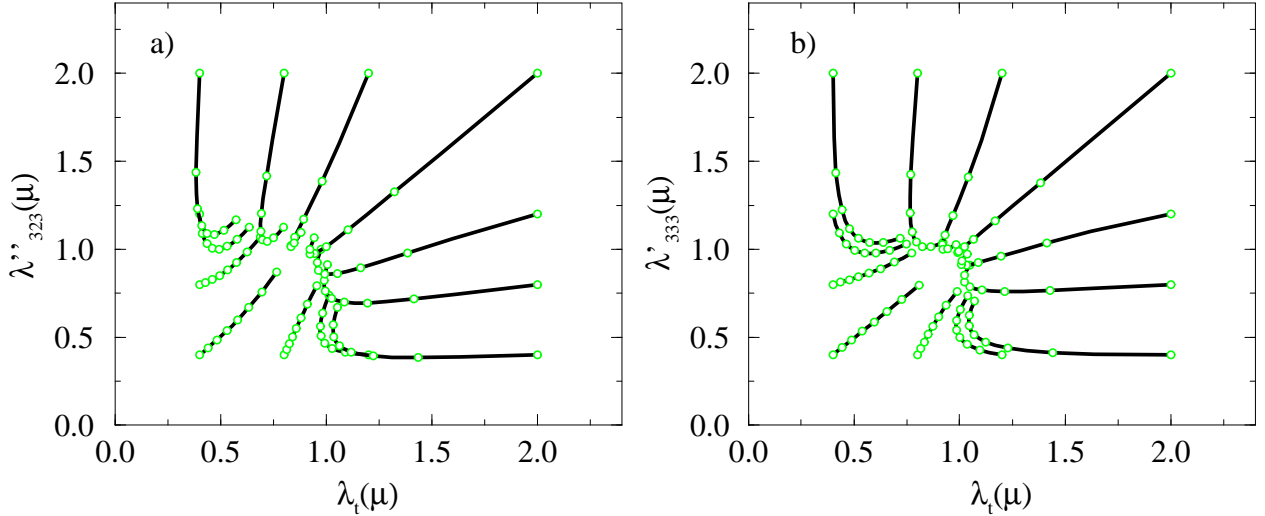}}
\end{center}
\caption{Running of $\lambda_t(\mu)$ and (a) $\lam''_{323}(\mu)$, (b)
$\lambda'_{333}(\mu)$. The circles are
separated by
two orders of magnitude of $\mu$/$\gev$. The flow is towards the region
(a) $\lambda_t \approx 1, {\lambda''}_{323} \approx 1$ and (b)
$\lambda_t \approx 1, {\lambda'}_{333} \approx 1$
for decreasing $\mu$.}
\label{fig:quasi}
\end{figure} 

Fig.~\ref{fig:fixed} only shows information on the quasi-fixed point
structure of $\lambda_t$. One would like to know if the \rpv\
couplings also exhibit the QFP behaviour when they are large.
Fig.~\ref{fig:quasi}a shows the running of both $\lt(\mu)$ and 
$\lam''_{323}(\mu)$ with the renormalisation scale $\mu$. 
$\lb$, $\ltau$ have been
switched off in this calculation. The two-loop \rpv-MSSM RGEs were run
through 14 orders of magnitude (roughly corresponding to running from
$M_U$ to $m_t$), using $\alpha_U=0.041$.  Although the figure shows
both couplings running toward 1 in the infra-red regime, it is clear
that it is difficult to make this statement quantitatively accurate.
The same conclusion holds for large $\lam'_{333}\not=0$ and $\lt$, as
shown in Fig.~\ref{fig:quasi}b. $\lam_{323}$ exhibits even less
focussing behaviour because it is not directly affected by QCD
interactions.

\subsubsection{Perturbative Limits \label{sec:pert}}
In Fig.~\ref{fig:fixed}b, we show the limits from perturbativity upon the
\rpv-couplings defined at $m_t$. We use a degenerate effective SUSY spectrum
at $m_t$ in this calculation but no finite SUSY threshold effects are
included. We include $\ltau$, $\lb$ however, because they make a large
difference to any Landau poles of Yukawa couplings at high $\tan
\beta$.  When we switch an \rpv\ coupling on, the curves in the figure
map out what value of the coupling is required at $m_t$ to produce a
value of 5 for at least one of the Yukawa couplings at $M_U$. In
practice, to good accuracy the point where a Yukawa coupling is 5 is
very close to the Landau pole of that coupling. The deviation between
the two curves shows the significant weakening effect of including
two-loop terms in the RGEs: 5$\%$, 12$\%$ and 10$\%$ for $\lam_{323}$,
$\lam'_{323}$ and $\lam''_{323}$ at high $\tan \beta$
respectively. The upper bound calculated in this way exhibits a strong
$\tan\beta$ dependence due to $\lb$, $\ltau$ contributing to the
running at high $\tan \beta$. Our one-loop bounds in
Fig~\ref{fig:fixed}b agree with the previous limits upon
$\lam''_{323}$ from perturbativity bounds provided in by Brahmachari
and Roy~\cite{prob}\footnote{Note that Brahmachari and Roy differ by a
factor of two in their convention for the $\lam''$ superpotential
terms.} to within 3\%.  One-loop bounds upon other $\lam''_{ijk}$ were
obtained by Goity and Sher~\cite{prob}.

\nsect{Conclusions \label{sec:conc}}
We have argued that \rpv\ is theoretically on equal footing with conserved
$R_p$. Since it can be realized in grand unified theories it is relevant for
unification. We then first determined the complete two-loop renormalisation
group equations for the dimensionless couplings of the unbroken supersymmetric
Standard Model. It is only at two-loop that Yukawa couplings affect the running
of the gauge coupling constants. We then considered three models of \rpv. We
have added to the MSSM in turn the three Yukawa operators $L_3L_2\Ebar_3$, $L_3
Q_3\Dbar_3$, and $\Ubar_3\Dbar_2\Dbar_3$. We considered their effects on
various aspects of the perturbative unification scenario. We have focused on
qualitative effects. A detailed search for a preferred model is beyond the
scope of this paper. We found several important effects. The unification scale
is shifted by up to $\pm20\%$. This is comparable to some threshold effects but
insufficient for string unification. $\alpha_s(M_Z)$ can be changed at most by
$\pm5\%$. The reduction which is favoured by the data is obtained close to the
perturbative limits of ${\lam'}_{333}$ and $\lam''_{323}$. We have obtained
the two-loop limit from perturbative unification for all three operators. 
For $\lam_{323}$ it is
equivalent to the laboratory bound for a slepton mass of $1.5\tev$ and for
$\lam'_{333}, \lam''_{323}$ is competitive for masses below $1\tev$. 
Two-loop limits from perturbativity are $5-12\%$ weaker than the one-loop
limit previously obtained. 
This is all
quite remarkable.
The \rpv-couplings can have significant effects on the entire Yukawa
unification picture. 
For bottom-tau unification we have found significant affects. For
$\lam_{323}\not=0$ bottom-tau unification could be obtained for values of $\tan
\beta<1.5$ were it not for the fact that the perturbative limit is reached.
For $\lam''_{323},\lam'_{333}\not=0$, 
we found new points of bottom-tau
unification at $\tan\beta=5-26$. 
Thus for $\tan \beta<30$, bottom-tau unification doesn't necessarily
correspond to top IR quasi fixed-point 
structure, as is the case in the MSSM \cite{ohman,ir,topfix,recentfp}. 
Indeed, the quasi-fixed structure is changed resulting in lower values of the
$\lam_t(m_t)$ prediction for $\lam'_{333}, \lam''_{323}$. This allows $\tan
\beta$ to increase, resulting in higher masses for the lightest CP-even Higgs,
relaxing severe constraints on the QFP.

\section*{Acknowledgements}
We thank Stefan Pokorski, Vernon Barger, Probir Roy and Marc Sher for helpful
conversations. This work was partially supported by PPARC. A.D acknowledges
the financial support from the Marie Curie Research Training Grant
ERB-FMBI-CT98-3438.

\nappend{Appendix}
We consider a group $G$ with representation matrices ${\bf t}^A\equiv
({\bf t})^{Aj}_i$. Then the quadratic Casimir $C(R)$ of a representation
$R$ is defined by
\beq
({\bf t}^A{\bf t}^A)^j_i= C(R) \delta^j_i.
\eeq
For $SU(3)$ triplets $q$ and for $SU(2)_L$ doublets $L$ we have
\beq
C_{SU(3)}(q)=\frac{4}{3}, \quad C_{SU(2)}(L)=\frac{3}{4}.
\eeq
For $U(1)_Y$ we have
\beq
C(f)=\frac{3}{5} Y^2(f),
\eeq
where $Y(f)$ is the hypercharge of the field $f$. The factor $3/5$
is the grand unified normalisation.

For the adjoint representation of the group
of dimension $d(G)$ we have
\beq
C(G)\delta^{AB}=f^{ACD}f^{BCD},
\eeq
where $f^{ABC}$ are the structure constants. Specifically for the
groups we investigate
\beq
C(SU(3)_C)=3,\quad C(SU(2)_L)=2,\quad C(U(1)_Y)= 0,
\eeq
and $C(SU(N))=N$. The Dynkin index is defined by
\beq
\mbox{Tr}_R({\bf t}^A{\bf t}^B)\equiv S(R)\delta^{AB}.
\eeq
For the respective fundamental representations $f$ we obtain
\barr
SU(3),\, SU(2):\quad &&S(f)=\half,\\
U(1)_Y:\quad&&S(f)=\frac{3}{5}Y^2(f),
\earr
where we have inserted the GUT normalisation for $U(1)_Y$.

The coefficients in the two-loop running of the gauge couplings
\eq{gaugerg} are given by \cite{bjorkjones}
\barr
B_a^{(1)}&=&(\frac{33}{5},1,-3),\\
B_{ab}^{(2)}&=&\left( \begin{array}{ccc}
199/25 & 27/5 & 88/5 \\
9/5 & 25 & 24 \\
11/5 & 9 & 14
\end{array}
\right),\\
C_{a}^{u,d,e}&=&\left( \begin{array}{ccc}
26/5 & 14/5 & 18/5 \\
6 & 6 & 2 \\
4 & 4 & 0
\end{array}
\right).
\earr

\end{document}